\documentclass[useAMS,usenatbib]{mn2e}

\usepackage[english]{babel}
\usepackage[latin1]{inputenc}
\usepackage{amsmath}
\usepackage{amsfonts}
\usepackage{amssymb}
\usepackage[]{graphicx}
\usepackage{verbatim}
\usepackage{mathrsfs}
\usepackage{enumerate} 
\usepackage{paralist}
\usepackage[small, bf]{caption}

\newif\ifAMStwofonts
\AMStwofontstrue

\title[Chemical evolution of classical and ultra-faint dwarf spheroidal galaxies]{Chemical evolution of classical and ultra-faint dwarf spheroidal galaxies}
\author[F. Vincenzo et al.]{F. Vincenzo$^{1}$\thanks{E-mail:
vincenzo@oats.inaf.it; matteucc@oats.inaf.it}, F. Matteucci$^{1,2,3}$\footnotemark[1], S. Vattakunnel$^{1,2}$ and G. A. Lanfranchi$^{4}$
\\
$^{1}$Dipartimento di Fisica, Sezione di Astronomia, Universit\`a di Trieste, via G.B. Tiepolo 11, 34100, Trieste, Italy\\
$^{2}$INAF Osservatorio Astronomico di Trieste, via G.B. Tiepolo 11, 34100, Trieste, Italy\\
$^{3}$INFN, Sezione di Trieste, Via Valerio 2, 34100, Trieste, Italy\\
$^{4}$N\'ucleo de Astrof\'isica Te\'orica, Universidade Cruzeiro do Sul - Rua Galv\~ao Bueno 868, CEP 01506-000, S\~ao Paulo, Brazil}

\begin{document}

\date{Accepted 2014 April 8.  Received 2014 April 1; in original form 2014 February 11.}

\pagerange{\pageref{firstpage}--\pageref{lastpage}} \pubyear{2014}

\maketitle

\label{firstpage}

\begin{abstract}

We present updated chemical evolution models of two dwarf spheroidal galaxies (Sculptor and Carina) and the first detailed chemical evolution models of two ultra-faint dwarfs (Hercules and Bo\"otes I). Our results suggest that the dwarf spheroidals evolve with a low efficiency of star formation, confirming previous results, and the ultra-faint dwarfs with an even lower one. Under these assumptions, we can reproduce the stellar metallicity 
distribution function, the $[\alpha/Fe]$ vs. $[Fe/H]$ abundance patterns and the total stellar and gas masses observed at the present time in these objects. In particular, for the ultra-faint dwarfs we assume a strong initial burst of star formation, with the mass of the system being already in place at early times. On the other hand, for the classical dwarf spheroidals the agreement with the data is found by 
assuming the star formation histories suggested by the Color-Magnitude diagrams and a longer time-scale of formation via gas infall. We find that all these galaxies should experience galactic winds, starting in all cases before $1$ Gyr from the beginning of their evolution. From comparison with Galaxy data, we conclude that it is unlikely that the ultra-faint dwarfs have been the building blocks of the whole Galactic halo, although more data are necessary before drawing firm conclusions.
\end{abstract}


\section{Introduction}

Orbiting around the Milky Way, there is a large number of satellite galaxies, most of which have so low average surface brightnesses and small effective radii that their detection was very difficult in the past: from 1937 up to 1994, only nine of them 
were discovered and that number remained unchanged until 2005. They are the so-called \textit{classical dwarf spheroidal galaxies} (dSphs), which are among the least luminous and most dark matter (DM) dominated galaxies which are observed today in the 
Universe. Dwarf spheroidal galaxies are classified as early-type since they are observed to possess very low gas mass at the present time and their stars are very iron-poor when compared to the Sun (see \citealt{tolstoy2009} and \citealt{koch2009} for an 
exhaustive review).

\par Color-magnitude diagram (CMD) fitting analysis revealed star formation rate (SFR) in dSphs to have been either continuous for a long time or occurring in bursts.  All dSph galaxies host an underlying very old stellar population with age 
$\ga10$ Gyr (e.g. \citealt{grebel1997}), and some of them are dominated by an intermediate-age stellar population with age in the range $4-8$ Gyr \citep{dall'ora2003}. Very few dSphs have been observed to host younger stars, which populate the so-called ``blue plume'' in the CMD, sign of a relatively recent star formation activity, which occurred up to $\sim2-3$ Gyr ago \citep{monelli2003}.

\par All such features led cosmologists to hypothesize dSphs to be the evolved small progenitor systems which merged in the past to form the actual large structures in the Universe as the stellar halo of the Milky Way, in the framework of the $\Lambda$CDM 
standard cosmological model \citep{helmi1999,bullock2001,harding2001,bullock2005}. However, successive deeper investigations revealed them not to possess all the right properties to be the aforementioned hypothetical survived progenitors of the Galactic 
halo (see \citealt{helmi2006} and \citealt{catelan2009} and references therein).

\par The Sloan Digital Sky Survey (SDSS) allowed in the past few years (from 2005 up to the present time) to discover a large number of new dwarf galaxies orbiting around the Milky Way with physical properties very similar to dSph galaxies but average surface 
brightnesses and effective radii even much smaller (see, e.g., \citealt{belokurov2007,belokurov2010}): for such reasons, they have been named as \textit{ultra-faint dwarf spheroidal galaxies} (UfDs). That is a mere naming convention, without any 
real physical motivation, since UfDs extend to fainter magnitudes and lower masses the same observed physical properties of dSphs. In fact UfD galaxies do not contain any gas at the present time and their stars are on average very iron-poor, with 
age $\ga10-12$ Gyr \citep{okamoto2012,brown2013}. So UfDs do not show any recent star formation activity. Such stellar systems soon aroused great interest in the scientific community, either for their extreme observed characteristics, or because such 
characteristics might shed light on the conditions of the Universe in the first billion years of its evolution (see \citealt{belokurov2013} for a detailed discussion). 
The UfD environment constitutes the best candidate for verifying whether a first  population of very massive and extremely metal-poor stars (the so-called Population III) might have existed or not. Such stars, at their death, should 
have enriched the interstellar medium (ISM) with some metals and therefore they might have left a chemical signature in the stars which were born immediately after, as some of the UfD stars (\citealt{ferrara2012}). As \citet{salvadori2012} had envisaged, this hypothesis might be supported by the observation of carbon-enhanced metal-poor (CEMP) stars in some UfD galaxies \citep{norris2010a,gilmore2013}, which might be directly linked to the CEMP-Damped Ly-$\alpha$ systems observed in the spectra of quasars at high redshift \citep{cooke2011a,cooke2011b}. In this scenario, the latter might be the high-redshift unevolved counterparts of some of the UfD galaxies. %\citep{salvadori2012}.

\par Aim of this work is to study how dSph and UfD galaxies have evolved, by reconstructing - going back in time - the chemical enrichment history of their ISM, starting from the chemical abundances derived today in the atmospheres of their stars. We will 
adopt a detailed chemical evolution model which is able to follow the evolution of several chemical species (H, He, C, N, O, $\alpha$-elements, Fe-peak elements, s- and r- process elements). This model is based on that presented by \citet{lanfranchi2004} 
and then used also in the works of \citet{lanfranchi2006a}, \citet{lanfranchi2006b}, \citet{lanfranchi2007}, \citet{lanfranchi2008}, \citet{cescutti2008}, and \citet{lanfranchi2010}.
\par \citet{lanfranchi2004} modeled the chemical evolution of six dSphs of the Local Group including Sextans, Sculptor, Sagittarius, Draco, Ursa Minor and Carina. Their main conclusions were that dSphs suffered from very low star formation efficiency, which 
caused the iron pollution from Type Ia SNe to become important when the $[Fe/H]$ of galaxy ISM was still very low. In this manner, and by assuming also intense galactic winds, they were able to explain the observed decrease in the trends of $[\alpha/Fe]$ 
vs. $[Fe/H]$. Galactic winds prevent the galaxy to form stars soon after the onset of the outflow and this leads the stellar metallicity distribution function (MDF) in dSphs to be peaked towards low $[Fe/H]$ abundances, almost $1.5$ dex below the one of 
Milky Way disc in the solar neighborhood, in agreement with observations. 
\par As interesting papers appeared recently, \citet{romano2013} modeled the chemical evolution of Sculptor by means of a new approach in a full cosmological framework, whereas \citet{koch2012,koch2013} present a first chemical evolution model of the 
Hercules UfD, based on the same numerical code we use in this work.
\par Here we focus on the chemical evolution of Sculptor and Carina, for which newer data are available, but especially we model the evolution of UfDs: Hercules and Bo\"otes I. In Section \ref{section2} we describe the adopted chemical evolution model. In 
Section \ref{section3} the data sample is presented and in Section \ref{section4} the results are shown. Finally, in Section \ref{section5} some conclusions are drawn.

\section{Description of the model} \label{section2}

\subsection{Assumptions}

We assumed dSph and UfD galaxies to form by infall of primordial gas in a pre-existing DM halo.  The infall rate obeys to a decaying exponential law with very short typical time-scales and the so-called \textit{infall mass} represents the reservoir of the 
infalling gas from which the stars form. 

Thanks to the detailed treatment of the main physical processes involved in the chemical enrichment of the galaxy ISM, we were able to predict the trends of the chemical abundances of many elements relative to iron as a function of $[Fe/H]$\footnote{We 
will use the following notation for the stellar chemical abundances: $[X/Y]=\log_{10}(N_{X}/N_{Y})_{\star}-\log_{10}(N_{X}/N_{Y})_{\odot}$, where $N_{X}$ and $N_{Y}$ are the volume density number of the atoms of the species $X$ and $Y$ respectively.}, the 
MDF and the total stellar and gas masses at the present time. 
Other observational constraints are represented by the history of star formation (SFH), which may be characterized either by a single long episode or by several bursts of star formation, according to the CMD-fitting analysis; the total mass of the DM halo and, finally, the effective radius of the luminous baryonic component of the galaxy. We assumed, when we do not have any information, a diffuse dark matter halo for these galaxies, with $S=\frac{r_{L}}{r_{DM}}$ between $0.2$ and $0.4$, where $r_{DM}$ represents the core radius of each galaxy DM halo.

\par The main assumptions of the model, which is similar to that of \citet{lanfranchi2004}, are the following:

\begin{enumerate}

\item each galaxy is modelled as a one zone with instantaneous and complete mixing of gas within it;
\item stellar lifetimes are taken into account (no instantaneous recycling approximation);
\item nucleosynthesis prescriptions include the yields of \citet{iwamoto1999} for Type Ia SNe, the yields of \citet{woosley1995} (with the corrections suggested by \citealt{francois2004}) for massive stars, and the  metallicity-dependent yields of 
\citet{vandenhoek1997} for low and intermediate mass stars.
\end{enumerate}

\par The galaxy can lose gas through galactic winds, which develop when the thermal energy of the gas $E_{g}^{th}(t)$ equals the binding energy of the gas $E_{g}^{b}(t)$. The latter quantity $E_{g}^{b}(t)$ is calculated by means of assumptions concerning 
the presence and the distribution of the DM which, for dSphs and UfDs, is usually represented as a diffuse halo (see \citealt{bradamante1998}), whereas the thermal energy of gas $E_{g}^{th}(t)$ is primarily determined by the energy deposited by SN 
explosions into the ISM (see also for more details \citealt{yin2011} and references therein).

\subsection{Basic equations}

If  $M_{g,i}(t)$ is the gas mass in the form of an element $i$ at the time $t$ within the ISM, its temporal evolution is described by the following basic equation:
\begin{equation}
\dot{M}_{g,i}=-\psi(t)X_{i}(t)+R_{i}(t)+(\dot{M}_{g,i})_{inf}-(\dot{M}_{g,i})_{out}, \label{eq:dsph}
\end{equation}
where the quantity $ X_{i}(t)=M_{g,i}(t)/M_{gas}(t)$ represents the abundance by mass of the element $i$, with the summation over the abundances of all the elements in the gas mixture being equal to unity, and $M_{gas}(t)$ is the total gas mass of the 
galaxy at the time $t$. The functions $\phi(m)$ and $\psi(t)$ represent the initial mass function (IMF) and the star formation rate (SFR), respectively.
\begin{enumerate}
\item The first term in the right hand side of Eq.(\ref{eq:dsph}) represents the rate at which chemical elements are subtracted from the ISM by star formation activity. The SFR has the following simple form (Schmidt law with $k=1$, \citealt{schmidt1959}):
\begin{equation}
\psi(t)=\frac{dM_{gas}}{dt}=\nu M^{k}_{gas}, \label{SFR}
\end{equation}
where $\nu$ is the star formation efficiency ($[\nu]=Gyr^{-1}$), which represents the intensity of the star formation rate and we assume in this work $k=1$. The inverse of $\nu$ corresponds to the typical time-scale for the complete gas consumption if 
star formation were the only physical process affecting the ISM.

\item The second term in Eq.(\ref{eq:dsph}), $R_{i}(t)$, represents the returned mass in the form of an element $i$ that stars eject into the ISM per unit time through stellar winds and supernova explosions. This term contains all the prescriptions 
concerning stellar yields and supernova progenitor models.

\item The third term in Eq.(\ref{eq:dsph}), $(\dot{M}_{g,i})_{inf}$, is the rate at which the element $i$ is accreted during the initial infall event. The rate of gas infall is defined as:
\begin{equation}
(\dot{M}_{g,i})_{inf}=A \cdot X_{i,inf} \cdot e^{-t/\tau}, \label{infallrate}
\end{equation}
where $A$ is a suitable constant such that 
\begin{equation}
\sum_{i} \int_{0}^{t_{G}}{dt \; A \cdot X_{i,inf} \cdot e^{-t/\tau}}=M_{inf}, 
\end{equation}
with $M_{inf}$ being the infall mass and $\tau$ the so-called infall time-scale of the mass accretion. The gas out of which dSph and UfD galaxies form is assumed to be of primordial chemical composition, namely $X_{i,inf}=0$ for heavy elements. We assumed 
for dSphs an infall time-scale $\tau_{dSph}=0.5$ Gyr, while for UfDs we adopted $\tau_{UfD}=0.005$ Gyr, which is almost equivalent to assume the gas to be all present in the DM halo since the beginning. The reason for these choices will be clearer in 
the next sections and are dictated by fitting the observational constraints.

\item The last term in Eq.(\ref{eq:dsph}) represents the rate at which the element $i$ is lost through galactic winds. For each element $i$, the rate of gas loss at the time $t$  is assumed to be directly proportional to the SFR at the time $t$:
\begin{equation}
(\dot{M}_{g,i})_{out}=\omega_{i} \psi(t), \label{gasloss}
\end{equation}
where $\omega_{i}$ are free parameters describing the efficiency of the galactic wind, i.e. the intensity of the rate of gas loss in the form of the element $i$. Here is all the information regarding the energy released by SNe, as well the efficiency with 
which that energy is transformed into gas escape velocity. For dSph and UfD galaxies we supposed the galactic wind to be \textsl{normal}, namely with the $\omega$ parameter being equal for all the chemical elements, although we tested also models with 
different kind of \textsl{differential} winds, following \citet{marconi1994} and \citet{recchi2001}.
\end{enumerate}

\section{Data sample} \label{section3} 

In general, we have selected - when possible - only high-resolution data and we have scaled all of them to the same Solar abundances \citep{asplund2009}. The reason for this was to homogeneize the data as much as possible.

\subsubsection{Sculptor dSph}  \label{datasculptor}

Sculptor is a relatively faint ($M_{V}\approx-11.1$, from  \citealt{mateo1998}) stellar system, located at a distance of $79\pm4\;\mbox{kpc}$ \citep{koch2009} away from us. Its features make it a fair representative of a typical dSph galaxy of the Local 
Group and it is also one of the most extensively investigated. It hosts the most iron-poor stars ever observed in a dSph, with $[Fe/H]=-3.96\pm0.06$ dex \citep{tafelmeyer2010}.

\par We used the dataset of chemical abundances presented by \citet{shetrone2003}, \citet{tafelmeyer2010}, \citet{frebel2010} and \citet{kirby2012}. The data sample we used for the MDF is taken from \citet{romano2013}, who 
combined the data samples of the DART project \citep{battaglia2008a,starkenburg2010} and of \citet{kirby2009,kirby2010}.

\subsubsection{Carina dSph}

Carina is located at a distance of $94\pm5$ kpc and belongs to the faintest dSphs of the Local Group ($M_{V}\approx-9.3$, $\mu_{V}=25.5\pm0.4$ mag arcsec$^{-2}$, from \citealt{mateo1998}). In the dSph realm, Carina is almost unique, since its CMD reveals a bursty SFH, which gave rise to the today observed multiple stellar populations \citep{hurley-keller1998,tolstoy2003}.
\par The dataset of chemical abundances we used for the comparisons with the predictions of our models have been taken from the following works: \citet{shetrone2003}, \citet{koch2008}, \citet{lemasle2012} and \citet{venn2012}. The observed MDF has been 
taken from \citet{koch2006}.

\subsubsection{Hercules UfD}

The Hercules UfD was discovered by \citet{belokurov2007} from the analysis of Sloan Digital Sky Survey (SDSS) data. The line of sight toward Hercules is heavily contaminated by Galactic foreground stars since it resides at a lower Galactic latitude than 
the other UfDs \citep{dejong2008}. Furthermore, the mean radial velocity of Hercules stars is very similar to the mean radial velocity of thick disk stars.
\par Hercules appears clearly highly elongated, without any evidence of internal rotation ($\sigma_{v}\sim3.72\;$ km s$^{-1}$, from \citealt{aden2009b}). Such a large ellipticity with no rotational support might imply that Hercules is not in dynamical 
equilibrium and it is undergoing strong tidal distortions. Hercules lies at a distance of $132\pm12$ kpc from us, it has an absolute V-band magnitude $M_{V}=-6.6\pm0.3$ and a V-band surface brightness of only $\mu_{V}=27.2\pm0.6$ mag arcsec$^{-2}$ (see 
\citealt{aden2009b} and references therein).
\par The data sample we used for the chemical abundances in the Hercules UfD is taken from \citet{aden2011}, which provide the chemical abundances of calcium and iron, the only elements actually available for this galaxy. 

\begin{table*}
\begin{tabular}{c c c c c c c c c}
\hline
\multicolumn{9}{c}{\textbf{\normalsize Sculptor: model parameters}}\\
\hline
\hline
$\nu$ & $\omega$ & $\tau_{inf}$ & SFH & $M_{inf}$ & $M_{DM}$ & $r_{L}$ & $S=\frac{r_{L}}{r_{DM}}$ & IMF \\
\hline 
$[Gyr^{-1}]$ & & [Gyr] & [Gyr] & $M_{\odot}$ & $M_{\odot}$ & $[pc]$ & & \\
\hline
\hline
$0.01/0.05/0.1/0.2$ & $5/10/15/20$ & $0.5$ & $0-7$ & $1.0\cdot10^{8}$ & $3.4\cdot10^{8}$ & $260$ & $0.260$ & \citet{salpeter1955} \\
\hline
\end{tabular}
\caption[sculptorinput]{ {\textit{Table:} we summarize here the specific characteristics of all the chemical evolution models performed for the Sculptor dSph. \textit{Columns:} (1) star formation efficiency; (2) wind parameter; (3) infall time-scale; (4) 
period of major star formation activity \citep{deboer2012}; (5) total infall gas mass; (6) mass of the dark matter halo \citep{battaglia2008b}; (7) effective radius of the \textsl{luminous} (baryonic) matter \citep{walker2009}; (8) ratio between the core radius of the 
dark matter halo and the effective radius of the \textsl{luminous} matter; (9) initial mass function. }}
\label{sculptorinput}
\end{table*}

\begin{table*}
\begin{tabular}{l c | c c c c}
\hline
\multicolumn{6}{c}{\textbf{\normalsize Sculptor chemical evolution models}}\\
\hline
\hline
\multicolumn{2}{l} {Input parameters}  & \multicolumn{4}{l}{Model predictions} \\
\multicolumn{1}{c}{} & $\nu$ & $M_{\star,fin}$ & $M_{gas,fin}$  & $t_{wind}$ & $[Fe/H]_{peak}$ \\
\hline 
  & $[{Gyr}^{-1}]$ & $M_{\odot}$ & $M_{\odot}$ & [Gyr] & dex \\
\hline
\hline

$\omega=5$ & $0.01$ & $0.13\cdot10^{7}$ & $0.26\cdot10^{3}$ & $2.48$ & $-1.91$ \\
           & $0.05$ & $0.21\cdot10^{7}$ & $0.12\cdot10^{4}$ & $1.01$ & $-1.73$ \\
           & $0.1$ & $0.27\cdot10^{7}$ & $0.19\cdot10^{4}$ & $0.66$ & $-1.62$  \\
           & $0.2$ & $0.34\cdot10^{7}$ & $0.15\cdot10^{4}$ & $0.38$ & $-1.42$  \\
\hline
$\omega=10$ & $0.01$ & $0.12\cdot10^{7}$ & $0.14\cdot10^{3}$ & $2.48$ & $-1.93$ \\
           & $0.05$ & $0.18\cdot10^{7}$ & $0.67\cdot10^{3}$ & $1.01$ & $-1.79$ \\
           & $0.1$ & $0.21\cdot10^{7}$ & $0.69\cdot10^{3}$ & $0.66$ & $-1.71$ \\
           & $0.2$ & $0.24\cdot10^{7}$ & $0.29\cdot10^{3}$ & $0.38$ & $-1.60$ \\
\hline
$\omega=15$ & $0.01$ & $0.12\cdot10^{7}$ & $0.97\cdot10^{2}$ & $2.48$ & $-1.94$ \\
           & $0.05$ & $0.17\cdot10^{7}$ & $0.46\cdot10^{3}$ & $1.01$ & $-1.81$ \\
           & $0.1$ & $0.19\cdot10^{7}$ & $0.30\cdot10^{3}$ & $0.66$ & $-1.74$ \\
           & $0.2$ & $0.20\cdot10^{7}$ & $0.11\cdot10^{3}$ & $0.38$ & $-1.67$ \\
\hline
$\omega=20$ & $0.01$ & $0.12\cdot10^{7}$ & $0.75\cdot10^{2}$ & $2.48$ & $-1.94$ \\
           & $0.05$ & $0.17\cdot10^{7}$ & $0.35\cdot10^{3}$ & $1.01$ & $-1.81$ \\
           & $0.1$ & $0.17\cdot10^{7}$ & $0.12\cdot10^{3}$ & $0.66$ & $-1.76$ \\
           & $0.2$ & $0.18\cdot10^{7}$ & $0.58\cdot10^{2}$ & $0.38$ & $-1.71$ \\
\hline
\end{tabular}
%}
\caption[sculptortable]{ {\textit{Table:} we reported for each model its main predictions. \textit{Columns:} (1) wind parameter; (2) star formation efficiency; (3) predicted actual total stellar mass; (4) predicted actual total gas mass; (5) time of the 
onset of the galactic wind; (6) peak of the stellar MDF predicted by the models.}}
\label{sculptortable}
\end{table*}

\subsubsection{Bo\"otes I UfD}

The Bo\"otes I UfD was discovered by \citet{belokurov2006}, from the analysis of SDSS data. \citet{siegel2006} estimated the distance of Bo\"otes I by using 15 RR Lyrae variable stars as standard candles; they found a distance: $D=62\pm4$ kpc, very 
similar to that of the outermost halo globular clusters in our Galaxy. Finally, \citet{belokurov2006} reported an absolute magnitude $M_{V}=-5.8$ and an half-light radius $r_{h}\sim220$ pc.

\par The data samples of chemical abundances have been taken from the works of \citet{norris2010a} and \citet{gilmore2013} (``NY'' analysis, from \citealt{norris2010b}). The observed MDF has been worked out by \citet{lai2011}, by expanding their own sample to include non-overlapping stars from \citet{norris2010b} and \citet{feltzing2009}. The data sample of \citet{ishigaki2014}, not considered in this work for its lower resolution than the other studies, shares three stars in common with \citet{gilmore2013}; in any case, the two data samples turn out to agree with each other in the final chemical abundances of the $\alpha$-elements.

\section{Results} \label{section4}

\subsection{Chemical evolution of the Sculptor dSph} \label{sculptor_results}

\begin{figure}
\includegraphics[width=8cm]{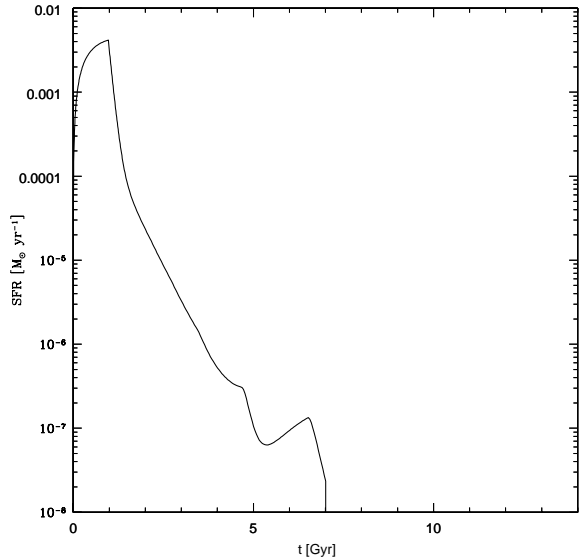}
     \caption{In this Figure we show the Sculptor SFR (in $M_{\odot}/yr$) as predicted by our model with $\nu=0.05$ Gyr$^{-1}$ and $\omega=10$, which turned out to well reproduce at the same time the observed abundance ratio patterns and the MDF, as well 
     as the other observed properties of the Sculptor dSph.}
     \label{sculptorSFR}
   \end{figure} 
   
\begin{figure}
\includegraphics[width=8cm]{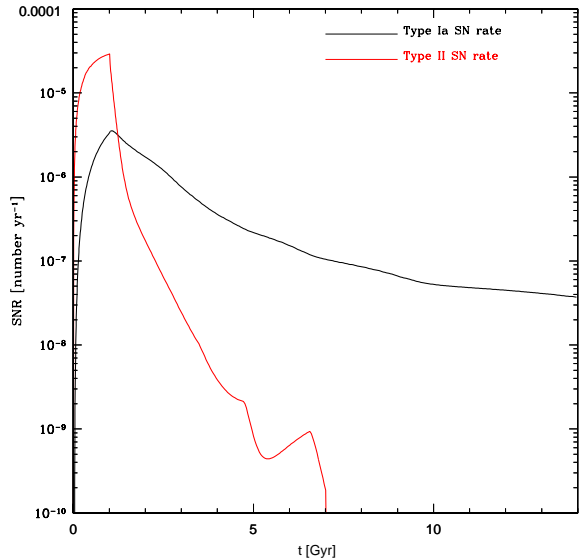}
     \caption{In this Figure we plot the Sculptor Type II and Type Ia supernova rates predicted by the same model as Fig.(\ref{sculptorSFR}).}
     \label{sculptorSNR}
   \end{figure}

\par We assumed the galaxy to form from the accretion of $1.0\cdot10^{8}\;M_{\odot}$ primordial gas, at an infall rate with a time-scale of $0.5$ Gyr. According to what has been inferred observationally by means of the CMD-fitting analysis 
\citep{deboer2012}, the SFH has been assumed to consist of only one long episode of star formation lasting $7$ Gyr. Following the observations, the galaxy is characterized by a massive and extended dark matter halo, with a mass of 
$3.4\cdot10^{8}\;M_{\odot}$ \citep{battaglia2008b} and we assumed a core radius $r_{DM}=1$ kpc. The effective radius of the \textsl{luminous} (baryonic) component of the galaxy has been set at the value of $260$ pc \citep{walker2009}. So the ratio between the core 
radius of the dark matter halo and the effective radius of the \textsl{luminous} matter is $S=\frac{r_{L}}{r_{DM}}=0.260$. We need these quantities in order to compute the binding energy of the gas and so the time at which the galaxy developed the 
galactic wind. The complete set of the parameters 
characterizing the chemical evolution models performed for the Sculptor dSph is summarized in Table \ref{sculptorinput}. We changed the values of $\nu$ (efficiency of star formation) and $\omega$ (wind parameter), in order to find a model able to 
reproduce the observed properties of the galaxy, such as the abundance ratios, the MDF and the present time gas and stellar masses. We tried models with $\omega=5,10,15$ and $20$; then, for each value of $\omega$, we varied the star formation efficiency 
$\nu$, setting it at the following values: $\nu=0.01,0.05,0.1,$ and $0.2\;Gyr^{-1}$.
   
\par In Fig.(\ref{sculptorSFR}) we show the trend of the predicted SFR as a function of time. By looking at this Figure, in the earliest stages of the galaxy evolution, the SFR increases very steeply because of the large amount of pristine gas accreted 
during the infall in a very short time-scale. In fact,  we have assumed the SFR, $\psi(t)$, to be proportional to the gas mass (see Eq.(\ref{SFR})). After the majority of the reservoir of gas $M_{inf}$ has been accreted, the star formation begins to 
decrease in time, since both SF activity and galactic wind subtract gas from the galaxy ISM.

\par In Fig.(\ref{sculptorSNR}) we compare the predicted Type II and Type Ia supernova rates. While SNe Ia occur on a large range of time-scales (from $\sim35$ Myr up to the age of the Universe), depending on the features of the progenitor system, the 
Type II SN rate traces very closely the trend of the SFR, because of the very short typical time-scales involved (from $\sim1$ up to $\sim35$ Myr). Thus, after the infall is substantially reduced, the Type II SN rate decreases much more steeply than the 
Type Ia SN rate. We predict a present time SNIa rate of $6.66\cdot10^{-4}\;SNuM$, where $SNuM$ represents the rate with units of one SN per $100\;yr$ per $10^{10}\;M_{\odot}$ stellar mass \citep{li2011}.

\par For each of the models we have calculated, Tab.(\ref{sculptortable}) reports its main predictions regarding the stellar and gas masses at the present time, the time of onset of the galactic wind, and the $[Fe/H]$-peak of the stellar MDF. In the first two columns, Tab.(\ref{sculptortable}) shows again the input parameters ($\nu$ and $\omega$) characterizing each model.

   \begin{figure*}    
    \includegraphics[width=14cm]{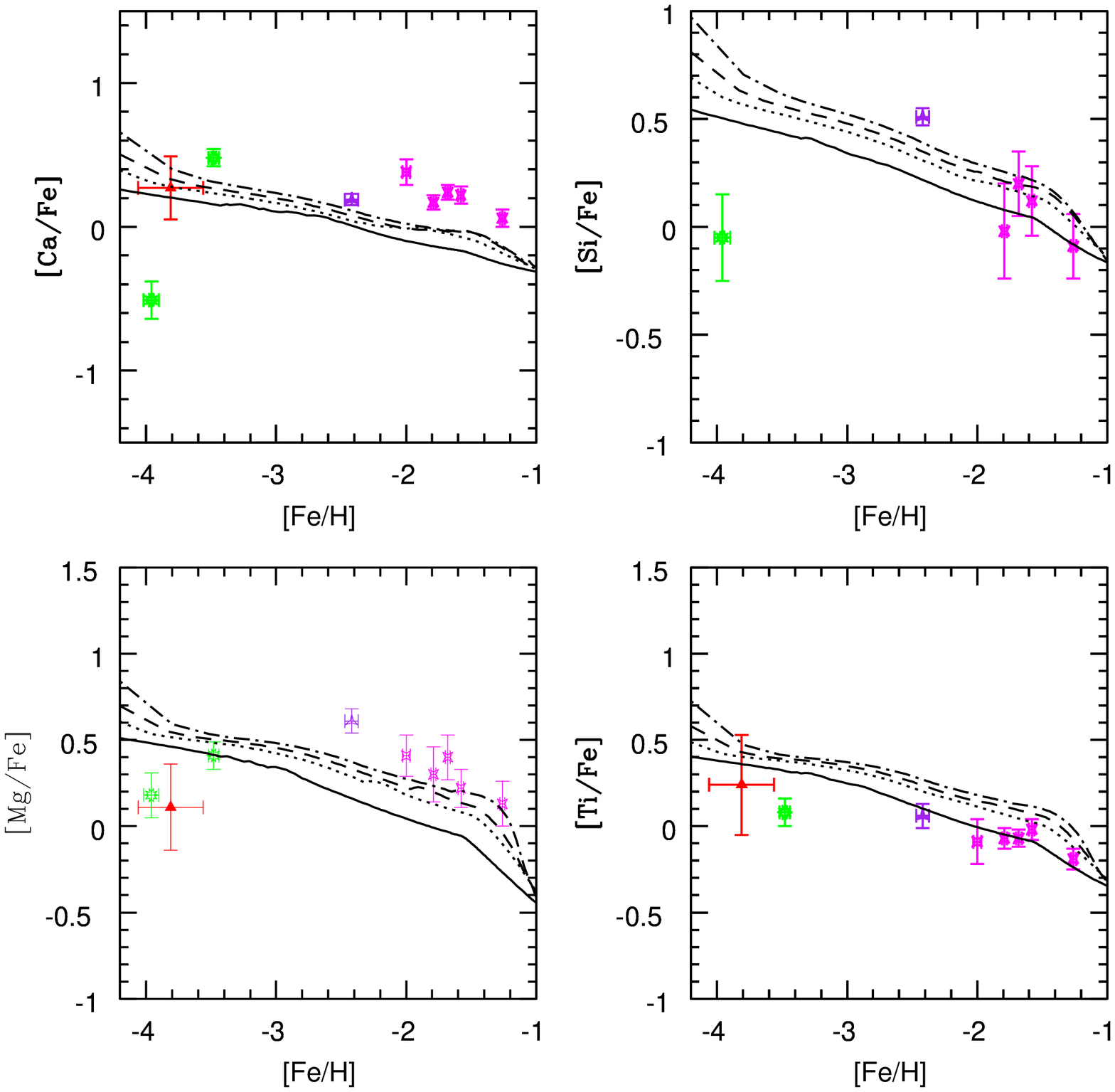} 
    \caption{
    In this Figure, we compare the observed Sculptor $[\alpha/Fe]$ vs. $[Fe/H]$ abundance patterns for calcium, silicon, titanium and magnesium with the trends predicted by our $\omega=10$ models. Data are from 
    \citet[in magenta]{shetrone2003}, \citet[in green]{tafelmeyer2010}, \citet[in red]{frebel2010} and \citet[in purple]{kirby2012}. The solid line corresponds to the model with $\nu=0.01\;Gyr^{-1}$, the dotted line to the model with $\nu=0.05\;Gyr^{-1}$, the long dashed line to the model with $\nu=0.1\;Gyr^{-1}$ and the dot-dashed line to the model with $\nu=0.2\;Gyr^{-1}$. Low star formation efficiencies $\nu$ cause to be small 
    the amount of iron coming from Type II SNe. So, when Type Ia SNe start to pollute the ISM with large amounts of iron, the $[Fe/H]$ of the ISM is very low. This gives rise to the decrease of $[\alpha/Fe]$ as $[Fe/H]$ increases. Furthermore, the onset of the galactic wind causes a steepening in the decrease of $[\alpha/Fe]$ as a function of $[Fe/H]$
    }
\label{alphasculptoromega10}
\end{figure*}

\par By looking at Tab.(\ref{sculptortable}), for a fixed value of the wind parameter $\omega$, models with increasing star formation efficiencies predict slightly larger final total stellar masses and, at the same time, slightly lower final total gas 
masses. This happens both because star formation is more efficient, and because the galactic wind is predicted to start earlier and earlier, being larger the number of SN events triggering the wind.

\par The wind parameter $\omega$ - which roughly represents the efficiency with which the chemical elements are carried out of the galaxy - characterizes the intensity of the outflow rate. By increasing the wind parameter $\omega$, our 
models predict the final total gas mass to be lower, since the galactic wind removes the interstellar gas from the galaxy potential well more efficiently. Moreover, as one would expect, the final total stellar mass slightly decreases if the 
efficiency of the galactic wind is strengthened.

 \begin{figure} 
 \centering   
    \includegraphics[width=9cm]{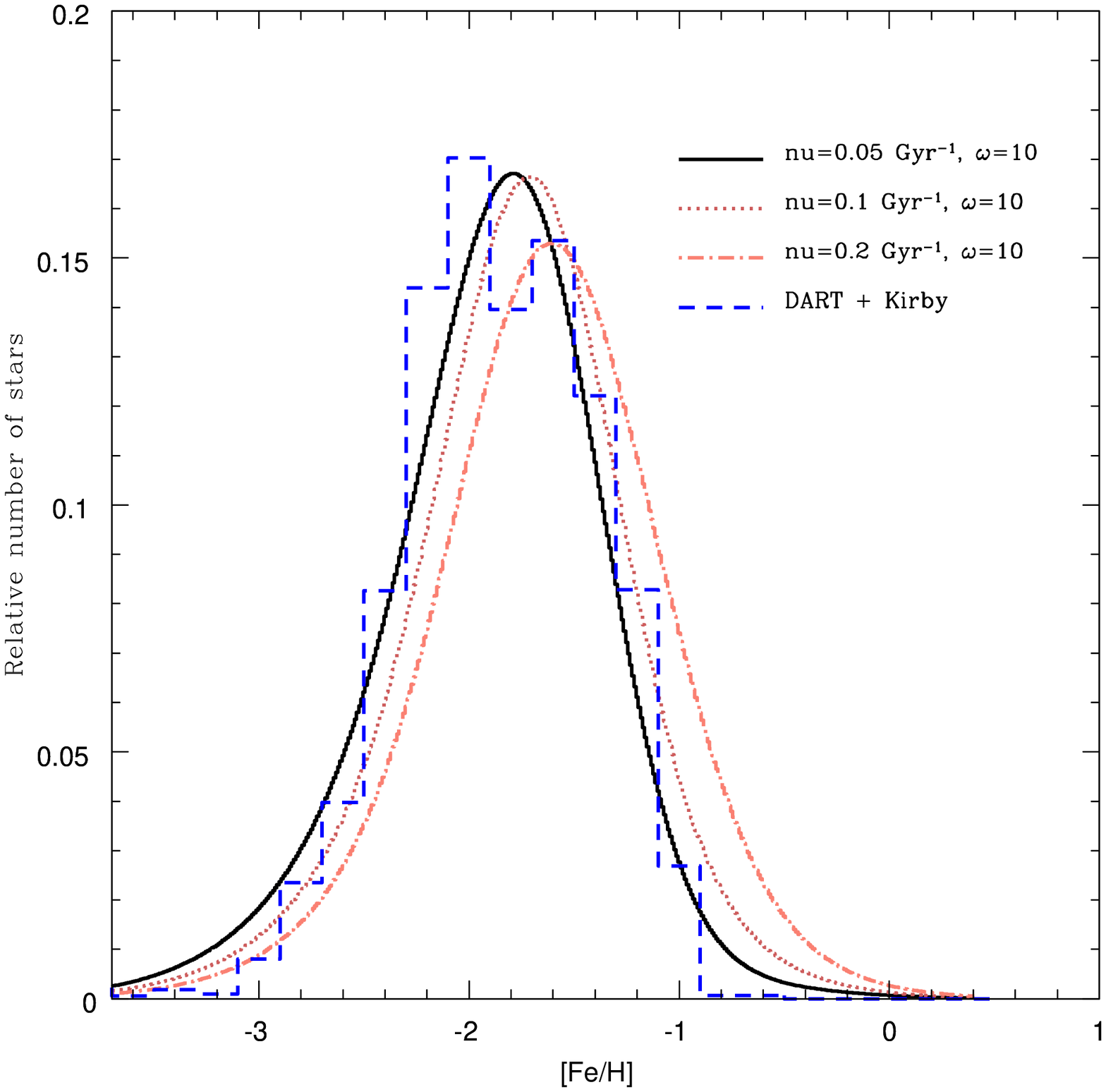} 
    \caption{
   The Figure shows the comparison between the observed stellar MDF of the Sculptor dSph and the predictions of our $\omega=10$ models. The observed dataset comes from \citet{romano2013}, who combined the data samples of the DART project 
   \citep{battaglia2008a,starkenburg2010} and of \citet{kirby2009,kirby2010}. By enhancing the star formation efficiency, the peak of the MDF shift towards higher $[Fe/H]$ abundance ratios, because of the shorter time-scales involved in the chemical 
   enrichment of the ISM. At the same time, the height of the peak diminishes, because of the stronger outflow rates ($\propto\psi(t)$), which clean up the galaxy of the gas; that is more evident by passing from $\nu=0.1\;Gyr^{-1}$ up to 
   $\nu=0.2\;Gyr^{-1}$. The best agreement is obtained with the $\nu=0.05$ Gyr$^{-1}$ and $\omega=10$ model.
    }
\label{mdfsculptoromega10}
\end{figure}

\par All the models at Tab.(\ref{sculptortable}) predict final stellar masses which are of the same order of magnitude as the observed one $M_{\star,obs}=(1.2\pm0.6)\cdot10^{6}\;M_{\odot}$, which has been derived by \citet{deboer2012} by integrating the 
SFR up to the present time; however, this mass is still quite uncertain, since it strongly depends not only on the IMF adopted, but also on the overall CMD-fitting technique. Moreover, all the models predict very low final gas masses 
($\sim10^{2}-10^{3}\;M_{\odot}$), which are lower than $M_{HI}\approx1.0\cdot10^{4}\;M_{\odot}$ estimated by \citet{mateo1998} and very much lower than the most recent estimate of \citet{grcevich2009}, which inferred for Sculptor an upper limit in the HI mass of 
$M_{HI}<2.34\cdot10^{5}\;M_{\odot}$, although affected by a very large uncertainty.

\par Finally, from Tab.(\ref{sculptortable}), it is possible to appreciate the net effect of changing the star formation efficiency $\nu$ and the wind parameter $\omega$ on the MDF-peak. The lowest $[Fe/H]$ at which the MDF has its peak is obtained by the 
models with the lowest $\nu$ and the highest $\omega$.

\subsubsection{Sculptor dSph: abundance ratios and interpretation} \label{sculptor_abundance}

As already mentioned, the $\alpha$-elements are mainly produced by Type II SNe, on short time-scales, whereas the bulk of iron and iron-peak elements ($\sim2/3$ of the total) are produced by Type Ia SNe on a very large range of typical time-scales. The 
residual quantities of iron and iron-peak elements ($\sim1/3$ of the total) are synthesized by Type II SNe. 
\par Different typical star formation time-scales can affect the $[\alpha/Fe]$ vs. $[Fe/H]$ diagram with a typical signature 
\citep{matteucci1990,matteucci2001}. In fact, low star formation efficiencies cause the amount of iron from Type II SNe to grow very slowly within the ISM. As the first Type Ia SNe start to explode, they deposit a large quantity of 
iron into the ISM and the $[\alpha/Fe]$ ratio is expected to steeply decrease at still very low $[Fe/H]$. Conversely, high star formation efficiencies give rise to large fractions of iron produced by Type II SNe. In this case, a plateau in the $[\alpha/Fe]$ vs. $[Fe/H]$ diagram is predicted and the decrease occurs at relatively high $[Fe/H]$ abundances (see \citealt{matteucci2001book}). Therefore, the abundance ratios in dSph reflect their low SFR \citep{lanfranchi2004}. In Fig.(\ref{alphasculptoromega10}) we show the effect of different $\nu$ values on the predicted $[\alpha/Fe]$ vs. $[Fe/H]$ relations. To lower further the SFR and thus steepening even more the decrease in the $[\alpha/Fe]$ vs. $[Fe/H]$, one can adopt higher values of the wind parameter $\omega$. In fact, the galactic wind subtracts gas to the SF and makes it to decrease. In the models of Fig.(\ref{alphasculptoromega10}), we assumed the outflow to remove all the chemical elements 
with the same  efficiency, i.e. with same $\omega$ parameter. 
\par We note that in Fig.(\ref{alphasculptoromega10}) the data relative to Mg lie below our model predictions. The reason for this can be uncertainties in the data or more likely uncertainties in the Mg yields. These anomalous low-Mg stars have been often associated with inhomogeneous enrichment on small scales from, e.g., single or at least few SNe \citep{koch2008,koch2008b,simon2010,frebel2012,keller2014}.  However, in small systems we should expect that different regions of the ISM mix on small typical timescales and, if the star formation timescales are large enough (and this is the case for dSphs and UfDs), the ISM should have had enough time to mix before new star formation occurs. So we should expect that the enrichment in low-mass and small-scale systems, with very low star formation efficiencies, is more homogeneous rather than more inhomogeneous. In fact, the cooling time for these small systems could be quite low (see \citealt{recchi2002}). Furthermore, dSphs seem to be supported by the 
random motions of stars rather than by regular motions. So, we expect that the thermalisation processes have been efficient and that the one-zone assumption is reasonable. Finally, small systems do not show evident abundance gradients and this is another indicator of a well mixed ISM.

\subsubsection{Sculptor dSph: MDF and interpretation} \label{sculptor_mdf}

One other important constraint to verify the goodness of a chemical evolution model is to check whether it is able to reproduce the observed stellar MDF. In Fig.(\ref{mdfsculptoromega10}), we show the comparison between the observed stellar MDF and our 
$\omega=10$ models. As shown in this Figure, by increasing the star formation efficiency, the predicted stellar MDF peaks towards higher $[Fe/H]$ abundances, since shorter time-scales are involved in the galactic chemical enrichment produced by 
successive generations of stars. At the same time, by increasing the $\nu$ parameter (especially from $\nu=0.1\;Gyr^{-1}$ to $\nu=0.2\;Gyr^{-1}$), the height of the MDF peak diminishes because of the stronger outflow rate ($\propto\psi(t)$) which leaves, 
on average, less gas available for star formation. The model which best reproduces the MDF shape of Sculptor is characterized by a star formation efficiency $\nu=0.05\;Gyr^{-1}$.

\subsubsection{Sculptor dSph: previous models} \label{sculptor_previous}

One of the first detailed chemical evolution models of Sculptor was presented by \citet{lanfranchi2004}. In this work we have adopted the same theoretical prescriptions and the same numerical code used in that first work

\par \citet{lanfranchi2004} did not possess yet an observed MDF dataset for the comparison with the models since only recently there have been obtained very precise determinations of the Sculptor SFH and physical characteristics, which have been used in 
our models as observational constraints.

\begin{table*}
\begin{tabular}{c c c c c c c c c}
\hline
\multicolumn{9}{c}{\textbf{\normalsize Carina: model parameters}}\\
\hline
\hline
$\nu$ & $\omega$ & $\tau_{inf}$ & SFH & $M_{inf}$ & $M_{DM}$ & $r_{L}$ & $S=\frac{r_{L}}{r_{DM}}$ & IMF \\
\hline 
$[Gyr^{-1}]$ & & [Gyr] & [Gyr] & $M_{\odot}$ & $M_{\odot}$ & $[pc]$ & & \\
\hline
\hline
$0.05/0.1/0.2$ & $5/10/20$ & $0.5$ & $0-2\,;\,2-4\,;\,7-9\,;\,9-11$ & $1.0\cdot10^{8}$ & $4.0\cdot10^{7}$ & $290$ & $0.3625$ & \citet{salpeter1955}\\
\hline
\end{tabular}
\caption[carinainput]{ {\textit{Table:} we summarize here the specific characteristics of all the chemical evolution models performed for the Carina dSph. \textit{Columns:} (1) star formation efficiency; (2) wind parameter; (3) infall time-scale; (4) period of major star formation activity \citep{rizzi2003}; (5) total infall gas mass; (6) mass of the dark matter halo \citep{gilmore2007}; (7) effective radius of the \textsl{luminous} (baryonic) matter \citep{gilmore2007}; (8) ratio between the core radius of the dark matter 
halo \citep{gilmore2007} and the effective radius of the \textsl{luminous} matter; (9) initial mass function. }}
\label{carinainput}
\end{table*}

\begin{table*}
\begin{tabular}{l c | c c c c}
\hline
\multicolumn{6}{c}{\textbf{\normalsize Carina  chemical evolution models}}\\
\hline
\hline
\multicolumn{2}{l} {Input parameters}  & \multicolumn{4}{l}{Model predictions} \\
\multicolumn{1}{c}{} & $\nu$ & $M_{\star,fin}$ & $M_{gas,fin}$  & $t_{wind}$ & $[Fe/H]_{peak}$ \\
\hline 
  & $[{Gyr}^{-1}]$ & $M_{\odot}$ & $M_{\odot}$ & [Gyr] & dex \\
\hline
\hline

$\omega=5$ & $0.05$ & $0.16\cdot10^{7}$ & $0.12\cdot10^{4}$ & $0.77$ & $-1.87$ \\
           & $0.1$ & $0.20\cdot10^{7}$ & $0.96\cdot10^{3}$ & $0.45$ & $-1.71$ \\
           & $0.2$ & $0.28\cdot10^{7}$ & $0.12\cdot10^{4}$ & $0.24$ & $-1.50$ \\
\hline
$\omega=10$ & $0.05$ & $0.13\cdot10^{7}$ & $0.65\cdot10^{3}$ & $0.77$ & $-1.90$ \\
            & $0.1$ & $0.14\cdot10^{7}$ & $0.18\cdot10^{3}$ & $0.45$ & $-1.88$ \\
            & $0.2$ & $0.17\cdot10^{7}$ & $0.24\cdot10^{3}$ & $0.24$ & $-1.55$ \\
\hline
$\omega=20$ & $0.05$ & $0.12\cdot10^{7}$ & $0.31\cdot10^{3}$ & $0.77$ & $-1.91$ \\
            & $0.1$ & $0.11\cdot10^{7}$ & $0.30\cdot10^{2}$ & $0.45$ & $-1.90$ \\
            & $0.2$ & $0.11\cdot10^{7}$ & $0.51\cdot10^{2}$ & $0.24$ & $-1.87$ \\
\hline
\end{tabular}
\caption[carinatable]{ {\textit{Table:} we reported for each model its main predictions. \textit{Columns:} (1) wind parameter; (2) star formation efficiency; (3) predicted actual total stellar mass; (4) predicted actual total gas mass; (5) time of the 
onset of the galactic wind; (6) peak of the stellar MDF predicted by the models.}}
\label{carinatable}
\end{table*}

\par However, although \citet{lanfranchi2004} had to work with very poorer dataset of chemical abundances, their predictions were similar to the results of this work. The main differences between the models calculated by this work and the ones of 
\citet{lanfranchi2004} are the following ones:

\begin{enumerate}

\item The galactic wind in \citet{lanfranchi2004} is	\textsl{differential}, i.e. some elements - in particular the products of SNe Ia - are lost more efficiently than others from the galaxy, following \citet{recchi2001}.

\item The models of \citet{lanfranchi2004} assume a more extended and massive dark matter halo ($M_{DM}=5.0\cdot10^{9}\;M_{\odot}$  and $S=\frac{r_{L}}{r_{DM}}=0.1$) than our models.

\item The infall gas mass in \citet{lanfranchi2004} is slightly higher than ours. In fact it is: $M_{inf}=5.0\cdot10^{8}\;M_{\odot}$.

\end{enumerate}

\citet{lanfranchi2004} suggested as best model for Sculptor the one with $\nu=0.2\;Gyr^{-1}$ and $\omega_{i}=13$. 

\subsection{Chemical evolution of the Carina dSph}

\begin{figure}
\includegraphics[width=8cm]{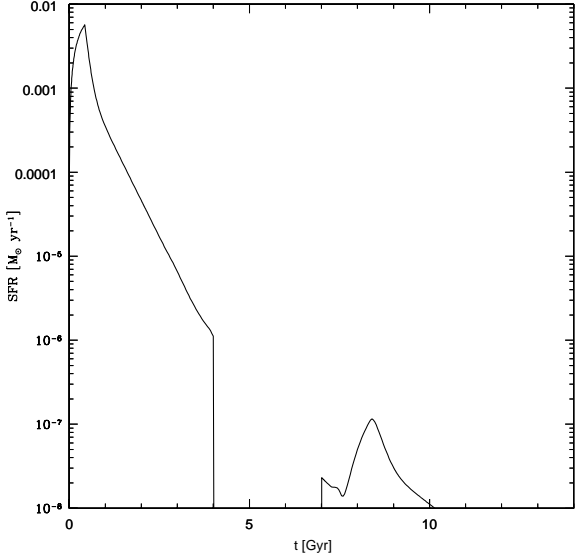}
     \caption{In this Figure we show the Carina SFR (in $M_{\odot}/yr$) as predicted by our model with $\nu=0.2$ Gyr$^{-1}$ and $\omega=10$.}
     \label{carinaSFR}
   \end{figure} 
   
\begin{figure}
\includegraphics[width=8cm]{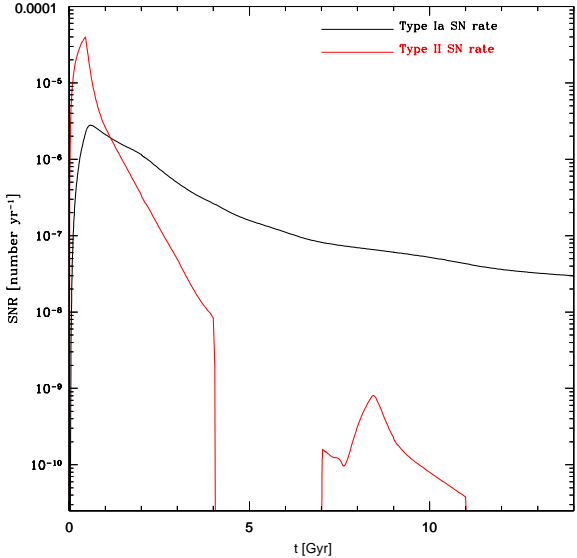}
     \caption{In this Figure we plot the Carina Type II and Type Ia supernova rates predicted by the same model as Fig.(\ref{carinaSFR}).}
     \label{carinaSNR}
   \end{figure} 

The most important difference between Sculptor and Carina resides in their SFH. While the Sculptor SFH is characterized by only one episode, with the SFR continuous for $7$ Gyr, the Carina dSph has been observed, from the analysis of its CMD, to have 
undergone a bursty SFH. So, following \citet{rizzi2003} and the suggestions of \citet{lanfranchi2006b}, we adopted for the Carina dSph a SFH characterized by $4$ episodes, all lasting $2$ Gyr, with the first two occurring from $0$ to $4$ Gyr and the 
second two occurring from $7$ to $11$ Gyr.

\par The galaxy DM halo is characterized by a total mass $M_{DM}=4.0\cdot10^{7}\;M_{\odot}$, and a core radius $r_{DM}=0.8\;kpc$ \citep{gilmore2007}. The effective radius of the \textsl{luminous} (baryonic) matter has been set at the value $r_{L}=290\;pc$ 
\citep{gilmore2007}. So, the ratio between the core radius of the DM halo and the effective radius of the baryonic matter is $S=\frac{r_{C}}{r_{L}}=0.3625$.

\par According to our model of chemical evolution, the galaxy assembled from the accretion of an \textsl{infall mass} $M_{inf}=1.0\cdot10^{8}\;M_{\odot}$, made up of primordial gas, with the infall time-scale being assumed to be $0.5$ Gyr. 

\par In Tab.(\ref{carinainput}), we have summarized all the chemical evolution models we have computed for the Carina dSph. Following the procedure used for the Sculptor dSph, we varied again only the $\nu$ and the $\omega$ parameters, namely the star 
formation efficiency and the wind parameter, respectively. 

\par In Fig.(\ref{carinaSFR}), it is shown the SFR of Carina dSph. By looking at that Figure, we can see that there is almost a continuous transition between the first ($0-2$ Gyr) and the second burst ($2-4$ Gyr), as well as from the third ($7-9$ Gyr) and the fourth burst ($9-11$ Gyr). This is due to the fact that the time interval between those episodes is so short that nor the galactic wind neither the gas restored by dying stars might have highly affected the gas mass and therefore the intensity of the SFR. In fact, the SFR has been assumed to be proportional to the gas mass in the galaxy. The discontinuity is clear only between the second ($2-4$ Gyr) and the third ($7-9$ Gyr) burst: during the temporary break ($4-7$ Gyr) of star formation, the gas mass in the galaxy decreases because of the strong outflow rate. So, at the epoch of the third burst (at $7$ Gyr), the intensity of the SFR is much lower than the intensity of the SFR at 
the end of the second burst (at $4$ Gyr), when there was much more gas. Here we did not consider external effects on the SFH (see for this \citealt{pasetto2011}).

    \begin{figure}

    \includegraphics[width=9cm]{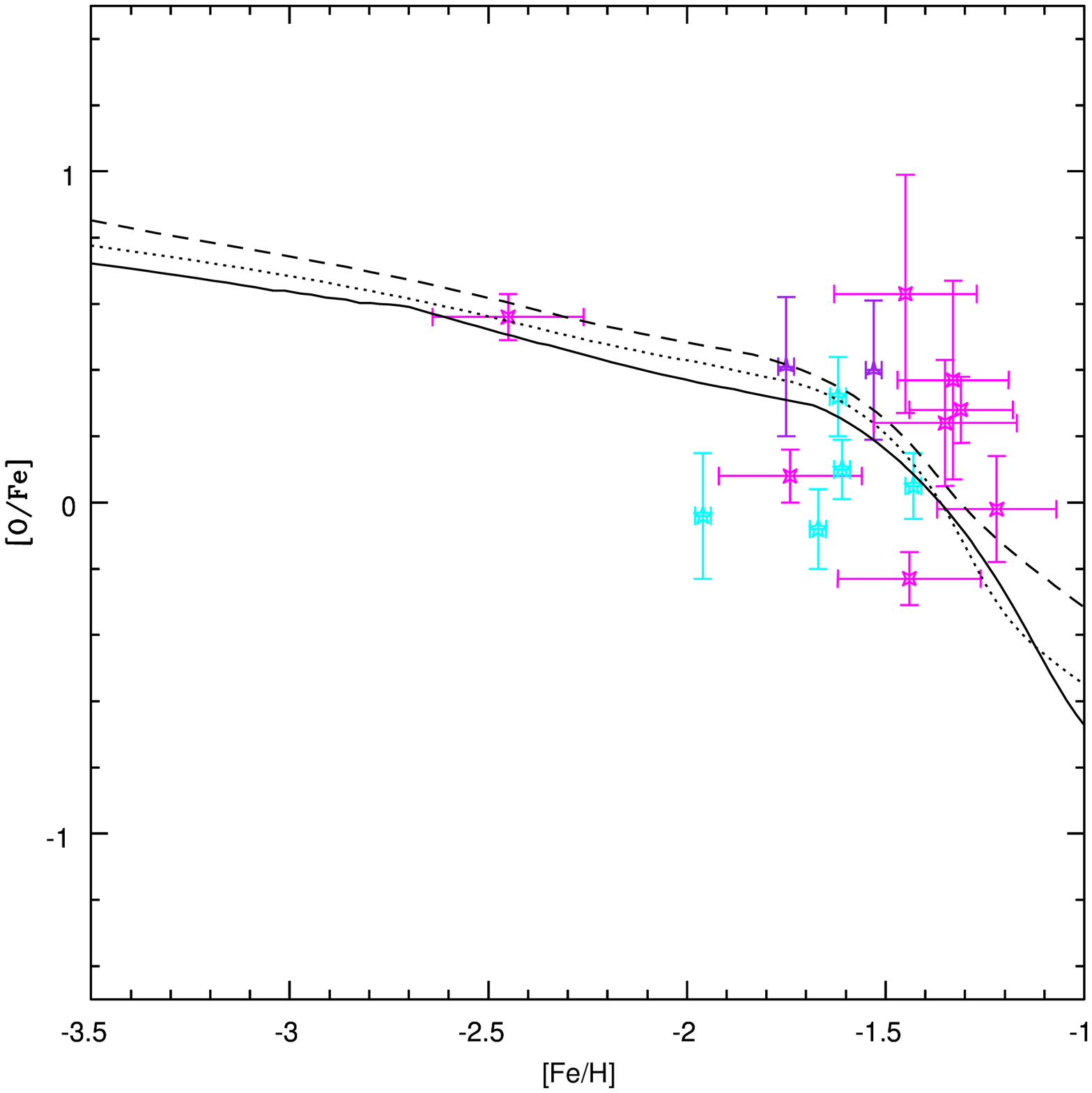} 
    \caption{
    In this Figure, we compare the observed $[O/Fe]$ vs. $[Fe/H]$ abundance ratios of Carina dSph member stars with the trends predicted by our $\omega=10$ models. The dataset consists of the data samples of \citet[in cyan]{shetrone2003}, \citet[in 
    magenta]{koch2008} and \citet[in purple]{venn2012}. The solid line corresponds to the model with $\nu=0.05\;Gyr^{-1}$, the dotted line to the model with $\nu=0.1\;Gyr^{-1}$ and the dashed line to the model with $\nu=0.2\;Gyr^{-1}$.}
    \label{carinaossigenoomega10}
    \end{figure}

    \begin{figure*}

    \includegraphics[width=14cm]{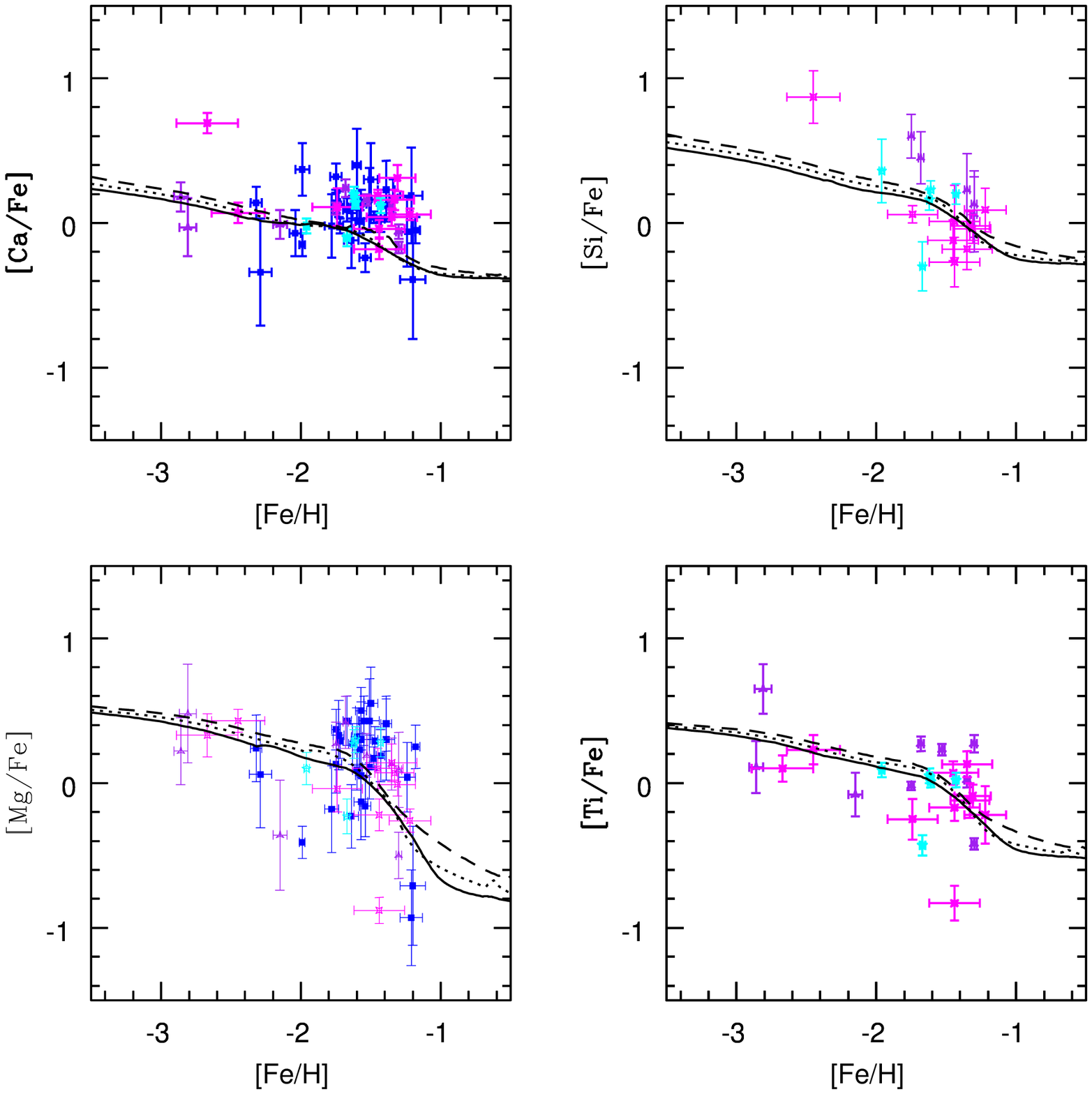} 
    \caption{
    In this Figure, we compare the $[\alpha/Fe]$ vs. $[Fe/H]$ abundance ratios for calcium, silicon, titanium and magnesium as observed in the Carina dSph member stars with the trends predicted by our $\omega=10$ models. The dataset consists of the data 
    samples of \citet[in cyan]{shetrone2003}, \citet[in magenta]{koch2008}, \citet[blue]{lemasle2012} and \citet[in purple]{venn2012}. The solid line corresponds to the model with $\nu=0.05\;Gyr^{-1}$, the dotted line to the model with $\nu=0.1\;Gyr^{-1}$ and the dashed line to the model with $\nu=0.2\;Gyr^{-1}$.
    }
\label{carinaalphaomega10}
\end{figure*} 

\par Fig.(\ref{carinaSNR}) shows the SN rates predicted by the best model. We predict a present time SNIa rate of $7.16\cdot10^{-4}\;SNuM$
\par In Tab.(\ref{carinatable}), we report the main predictions of all the chemical evolution models which we have performed for the Carina dSph. In the first two columns, we listed the input parameters ($\nu$ and $\omega$) characterizing each model. 
Briefly, by looking at the Table, the reader can note again that enhancing the star formation efficiency $\nu$, for a fixed value of the wind parameter $\omega$, will cause the peak of the MDF to shift towards higher values of $[Fe/H]$, the galactic wind 
to begin at earlier times, and the total stellar mass at the present time to be larger. Conversely, for a fixed value of the star formation efficiency $\nu$, models with higher values of the wind parameter $\omega$ predict the MDF to shift towards lower 
$[Fe/H]$, and both the stellar and gas masses to be lower at the present time. In order to explain such trends, the same arguments expressed in section \ref{sculptor_results} regarding the Sculptor dSph are also valid here, and we will not repeat them 
again. 
\par From Tab.(\ref{carinatable}), all the chemical evolution models we calculated predict both final gas and stellar masses at the present time which are in agreement with the observations ($M_{gas,obs}<2.1\cdot10^{2}\;M_{\odot}$ from 
\citealt{grcevich2009}, and $M_{\star,obs}\approx1.0\cdot10^{6}\;M_{\odot}$ from \citealt{dekel1986}).

\subsubsection{Carina dSph: abundance ratios and interpretation}

Since we adopted low star formation efficiencies, the $[\alpha/Fe]$ ratio is predicted to decrease at very low $[Fe/H]$, and such a decrease steepens even more once the galactic wind starts (see also section \ref{sculptor_abundance}). These trends are 
illustrated in Fig.(\ref{carinaossigenoomega10}) for the pattern of $[O/Fe]$ vs. $[Fe/H]$ abundance ratios and in Fig.(\ref{carinaalphaomega10}) for the calcium, silicon, magnesium and titanium abundance ratios. Especially for the oxygen, there is quite a good agreement between the predictions of our models and the 
observations.

    \begin{figure}
 %\centering   
    \includegraphics[width=9cm]{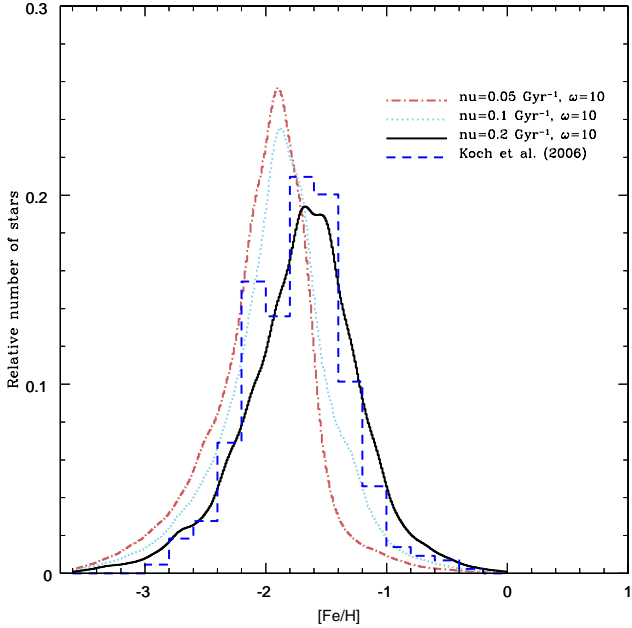} 
    \caption{
   The Figure shows the comparison between the observed stellar MDF in the Carina dSph (data from \citealt{koch2006}) and the stellar MDFs predicted by our $\omega=10$ models. The model which best reproduces the observed MDF is characterized by 
   $\nu=0.2\;Gyr^{-1}$ and $\omega=10$.
    }
\label{mdfcarinaomega10}
\end{figure} 

\subsubsection{Carina dSph: MDF and interpretation}

In Fig.(\ref{mdfcarinaomega10}), we compare the observed stellar MDF with the one predicted by our $\omega=10$ models. The best matching is given by the model with $\nu=0.2\;Gyr^{-1}$ and $\omega=10$. Therefore, star formation in Carina was bursty and 
characterized by a higher efficiency - and so by lower typical time-scales - than in Sculptor. The most evident effect caused by their different typical star formation time-scales resides in the $[Fe/H]$ of their MDF-peak, which is predicted by the best 
model for Sculptor ($\nu=0.05\;Gyr^{-1}$) to be at $[Fe/H]=-1.79$ dex, whereas the Carina best model predicts it at $[Fe/H]=-1.55$ dex.

\subsubsection{Carina dSph: previous models}

The first detailed chemical evolution model calculated for the Carina dSph was presented in \citet{lanfranchi2004} and its main basic differences from our models have been summarized in section \ref{sculptor_previous}. They assumed all dSphs to be 
characterized by the same infall mass and the same mass of DM halo. Conversely, they were able to vary the SFH from galaxy to galaxy, as it was given by the CMD-fitting analysis. They adopted for the Carina dSph a SFH taken from \citet{hernandez2000}, 
which was not confirmed by later and more accurate observations \citep{rizzi2003}. \citet{lanfranchi2006b} adopted the SFH from \citet{rizzi2003} and were able to reproduce the observed stellar MDF. 
\par The final results of both chemical evolution models of \citet{lanfranchi2004} and \citet{lanfranchi2006b} are very similar to the ones found by this work, although they adopted a \textsl{differential} galactic wind \citep{recchi2001}. The best model 
of \citet{lanfranchi2006b} has $\nu=0.14\;Gyr^{-1}$ and $\omega_{i}=5$.

\subsection{Chemical evolution of the Hercules UfD} \label{hercules:modelresults}

The chemical evolution models we ran for the Hercules UfD adopt the same theoretical prescriptions - and so the same numerical code - as the ones adopted for the dSph galaxies. The mass of the DM halo $M_{DM}$, the effective radius of 
the \textsl{luminous} (baryonic) component of the galaxy, as well as the SFH of the galaxy have been taken from the observations and kept fixed for all the models. The main difference resides in the fact that UfDs suffered an even slower SFR than dSphs 
(see also \citealt{salvadori2009}).

\begin{table*}
\begin{tabular}{c c c c c c c c c}
\hline
\multicolumn{9}{c}{\textbf{\normalsize Hercules: parameters of the models}}\\
\hline
\hline
$\nu$ & $\omega$ & $\tau_{inf}$ & SFH & $M_{inf}$ & $M_{DM}$ & $r_{L}$ & $S=\frac{r_{L}}{r_{DM}}$ & IMF \\
\hline 
$[Gyr^{-1}]$ & & [Gyr] & [Gyr] & $M_{\odot}$ & $M_{\odot}$ & $[pc]$ & & \\
\hline
\hline
$0.002/0.003/0.005/0.008$ & $10$ & $0.005$ & $0-1$ & $1.0/2.5/5.0\cdot10^{7}$ & $1.9\cdot10^{6}$ & $330$ & $0.3$ & \citet{salpeter1955} \\
\hline
\end{tabular}
\caption[herculesinput]{ {\textit{Table:} we summarize here the specific characteristics of all the chemical evolution models performed for the Bo\"otes I UfD. \textit{Columns:} (1) star formation efficiency; (2) wind parameter; (3) infall time-scale; (4) 
period of major star formation activity \citep{dejong2008,sand2009}; (5) total infall gas mass; (6) mass of the dark matter halo \citep{aden2009a}; (7) effective radius of the \textsl{luminous} (baryonic) matter \citep{martin2008}; (8) ratio between the core radius of the 
dark matter halo and the effective radius of the \textsl{luminous} matter; (7) initial mass function.}}
\label{herculesinput}
\end{table*}

\begin{table*}
\begin{tabular}{l c c | c c c c}
\hline
\multicolumn{7}{c}{\textbf{\normalsize Hercules chemical evolution models}}\\
\hline
\hline
\multicolumn{3}{l} {Input parameters}  & \multicolumn{4}{l}{Model predictions} \\
\multicolumn{1}{c}{} & $\nu$ & $\omega$ & $M_{\star,fin}$ & $M_{gas,fin}$  & $t_{wind}$ & $[Fe/H]_{peak}$ \\
\hline 
  & $[{Gyr}^{-1}]$ & & $M_{\odot}$ & $M_{\odot}$ & [Gyr] & dex \\
\hline
\hline
$M_{inf}=1.0\cdot10^{7}\;M_{\odot}$ & $0.002$ & $10$ & $0.11\cdot10^{5}$ & $0.35\cdot10^{1}$ & $0.97$ & $-3.06$ \\
           & $0.003$ & $10$ & $0.14\cdot10^{5}$ & $0.49\cdot10^{1}$ & $0.76$ & $-3.03$ \\
           & $0.005$ & $10$ & $0.17\cdot10^{5}$ & $0.18\cdot10^{1}$ & $0.56$ & $-2.99$ \\
           & $0.008$ & $10$ & $0.21\cdot10^{5}$ & $0.82\cdot10^{0}$ & $0.41$ & $-2.94$ \\
\hline
$M_{inf}=2.5\cdot10^{7}\;M_{\odot}$ & $0.002$ & $10$ & $0.28\cdot10^{5}$ & $0.88\cdot10^{1}$ & $2.41$ & $-3.05$ \\
           & $0.003$ & $10$ & $0.42\cdot10^{5}$ & $0.13\cdot10^{2}$ & $1.36$ & $-2.87$ \\
           & $0.005$ & $10$ & $0.66\cdot10^{5}$ & $0.22\cdot10^{2}$ & $0.92$ & $-2.67$ \\
           & $0.008$ & $10$ & $0.85\cdot10^{5}$ & $0.22\cdot10^{2}$ & $0.71$ & $-2.62$ \\
\hline
$M_{inf}=5.0\cdot10^{7}\;M_{\odot}$ & $0.002$ & $10$ & $0.55\cdot10^{5}$ & $0.48\cdot10^{8}$ & no wind & $-3.05$ \\
           & $0.003$ & $10$ & $0.85\cdot10^{5}$ & $0.26\cdot10^{2}$ & $6.76$ & $-2.87$ \\
           & $0.005$ & $10$ & $0.14\cdot10^{6}$ & $0.44\cdot10^{2}$ & $1.68$ & $-2.65$ \\
           & $0.008$ & $10$ & $0.22\cdot10^{6}$ & $0.71\cdot10^{2}$ & $1.07$ & $-2.43$ \\
\hline
\end{tabular}
\caption[herculestable]{ {\textit{Table:} we reported for each model with \textsl{normal wind} its main predictions. \textit{Columns:} (1) Infall mass; (2) star formation efficiency; (3) wind parameter; (4) predicted actual total stellar mass; (5) 
predicted actual total gas mass; (6) time of the onset of the galactic wind; (7) peak of the stellar MDF predicted by the models.}}
\label{herculestable}
\end{table*}

\par We assumed for the Hercules UfD a DM halo with mass $M_{DM}=1.9\cdot10^{6}\;M_{\odot}$ \citep{aden2009a}. The fraction between the core radius of the DM halo and the effective radius of the baryonic component of the galaxy has been set at 
$S=\frac{r_{L}}{r_{DM}}=0.3$, where the effective radius of the \textsl{luminous} (baryonic) component of the galaxy has been assumed to be $r_{L}=330\;pc$ \citep{martin2008}.
\par Since the CMD-fitting analysis derived a SFH concentrated in the very early epochs of the galaxy evolution \citep{dejong2008,sand2009}, we assumed the galaxy to have formed stars in the first Gyr of its evolution (see Tab.(\ref{herculesinput}). This 
is the only SFH which allowed us to reproduce all the observed features of the galaxy.

\par Tab.(\ref{herculesinput}) summarizes the parameters of the most relevant models we calculated for the chemical evolution of the Hercules UfD. We tested models with three different infall masses: $M_{inf}=1.0$, $2.5$, and $5.0\cdot10^{7}\;M_{\odot}$, 
which have been accreted inside the potential well of the dark matter halo in a very short infall time-scale: $\tau_{inf}=0.005$ Gyr. This is equivalent to assume that the gas mass is all present since the beginning. The chemical composition of the 
infalling gas has been assumed to be primordial. 

\begin{figure}
 
    \includegraphics[width=9cm]{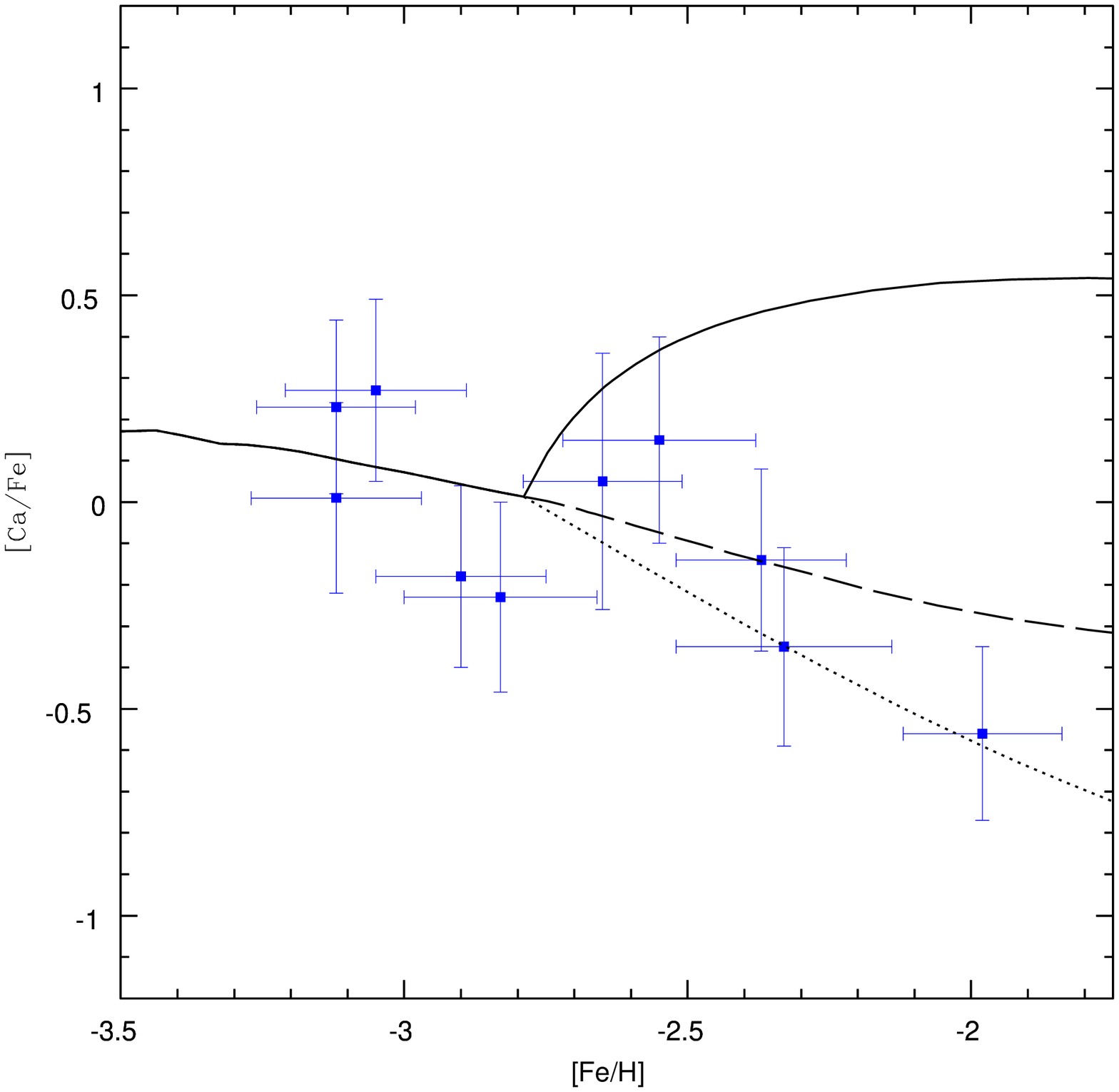} 
    \caption{
    In this Figure, we compare the $[Ca/Fe]$ vs. $[Fe/H]$ trends in Hercules predicted by models with different implemenentations of the galactic wind. The dataset has been taken from \citet{aden2011}. The curve in solid line corresponds to the model with 
    iron-enhanced differential wind \citep{recchi2001}; the dotted line corresponds to the prediction of the model with $\alpha$-enhanced differential wind \citep{marconi1994} , whereas the dashed line represents the prediction of the model with normal 
    wind. The observed dataset (blue squares with errorbars) are from \citet{aden2011}.
    }
\label{alphaherculesconfrontoventi}
\end{figure} 

\begin{figure}
    \includegraphics[width=9cm]{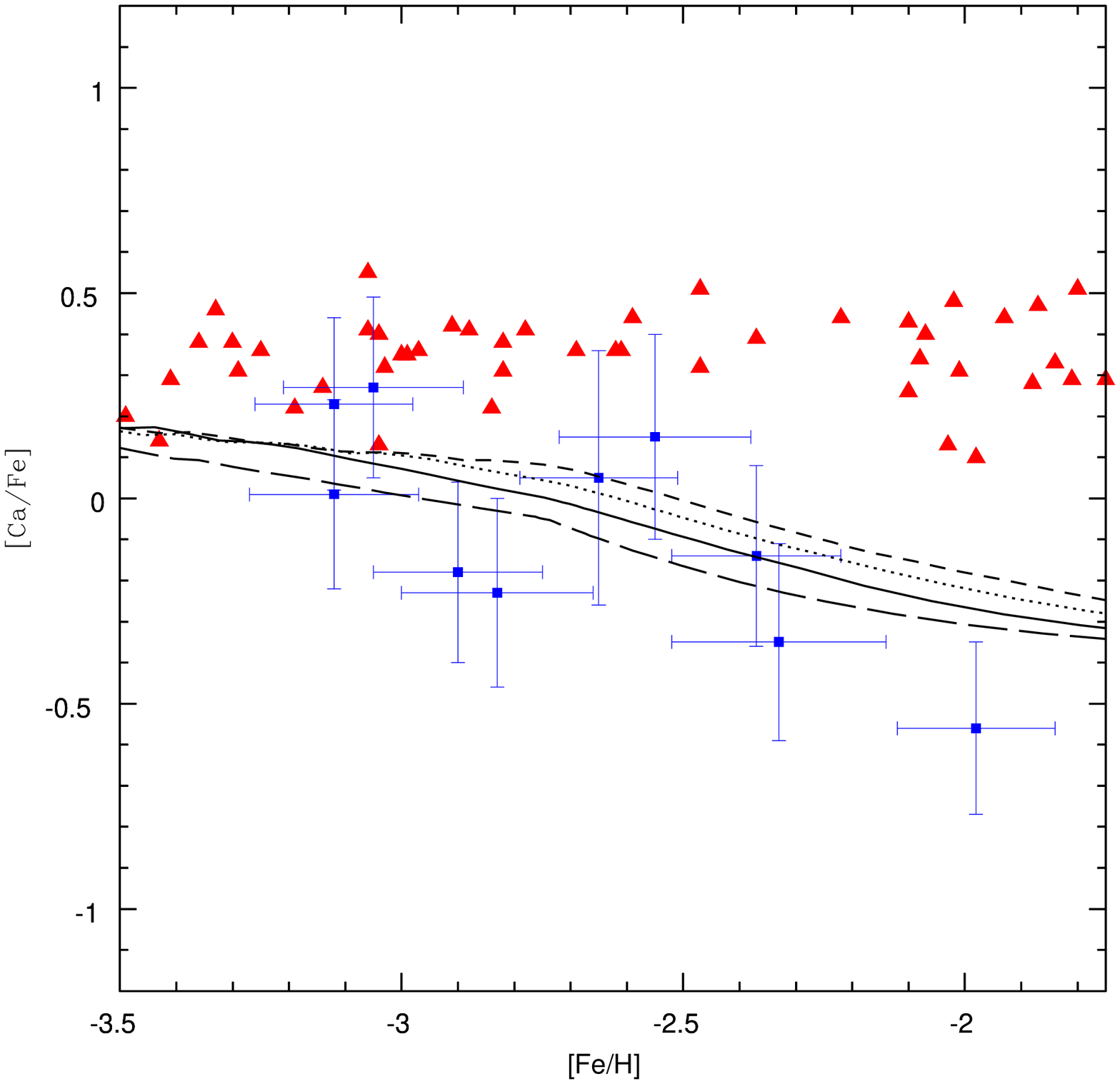} 
    \caption{
    In this Figure, we compare the $[Ca/Fe]$ vs. $[Fe/H]$ abundance pattern as observed in the Hercules UfD member stars (data from \citealt{aden2011}) with the predictions of our models with $\omega=10$ (normal wind) and 
    $M_{inf}=1.0\cdot10^{7}\;M_{\odot}$. We show also the abundance pattern of $[Ca/Fe]$ vs. $[Fe/H]$ as observed in Galactic halo stars (red triangles, data from \citealt{gratton2003,reddy2003,cayrel2004,reddy2006}). The prediction of the model with 
    $\nu=0.002\;Gyr^{-1}$ is in long dashed line; the model with $\nu=0.003\;Gyr^{-1}$ is in solid line;  the model with $\nu=0.005\;Gyr^{-1}$ is illustrated in dotted line, whereas the model with $\nu=0.008\;Gyr^{-1}$ is represented by the dashed line. 
    By looking at this Figure, it is clear that Hercules UfD member stars suggest different trend of $[Ca/Fe]$ as a function of $[Fe/H]$.
    }
\label{calciumhercules}
\end{figure} 

\begin{figure}
    \includegraphics[width=9cm]{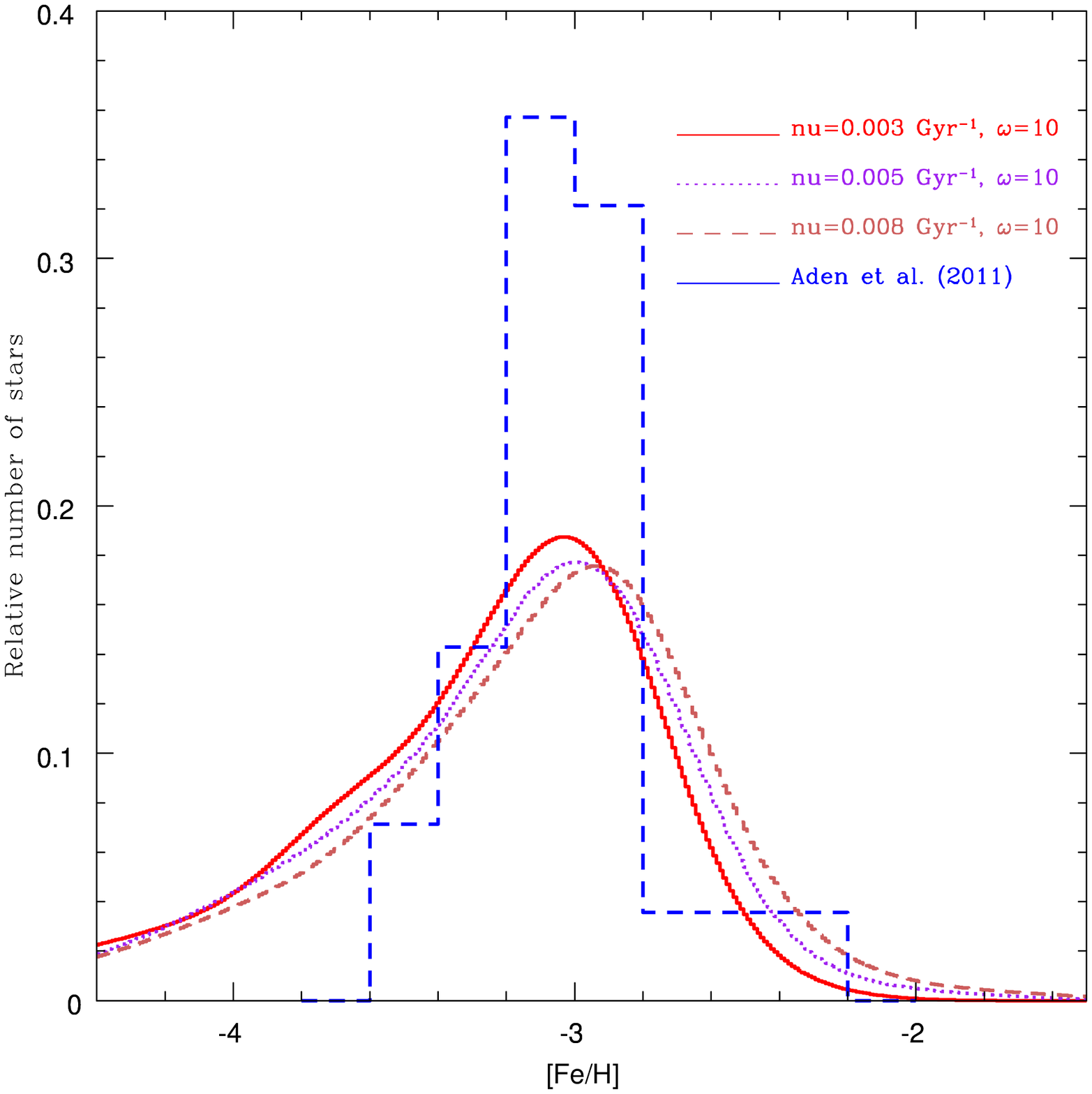} 
    \caption{
   In this Figure, we show the comparison between the observed stellar MDF of the Hercules UfD (data from \citealt{aden2011}) and the predictions of our models with $\omega=10$ (normal wind) and $M_{inf}=1.0\cdot10^{7}\;M_{\odot}$. The dataset is built up 
   on a sample of only $28$ RGB stars, with $[Fe/H]$ values in a very tight range, between $[Fe/H]=-3.1$ dex and $[Fe/H]=-2.9$ dex and very low relative numbers of stars in the other bins towards both lower and higher $[Fe/H]$ abundances. The MDF 
   predicted by our models are based on a large number of artificial stars, according to the SFH and to the IMF assumed. We are not able to reproduce the height of the observed-MDF peak. Our choice of the best model is oriented towards the model which 
   reproduces the $[Fe/H]_{peak}$ abundance of the observed MDF, which turns out to have $\nu=0.003\;Gyr^{-1}$ and $\omega=10$.}
\label{mdfhercules}
\end{figure} 

\par Three different implementations of the galactic wind were tested:
\begin{enumerate}

\item \textsl{normal wind}: all the chemical elements are characterized by the same wind parameter $\omega$. This means that the galactic wind expels all the elements with same efficiency.

\item \textsl{$\alpha$-enhanced differential wind} \citep{marconi1994}: the $\alpha$-elements are expelled by galactic winds with more efficiency than the iron. This means that $\omega_{\alpha}>\omega_{Fe}$. We implemented this galactic wind by adopting: 
$\omega_{\alpha}=1.0\cdot\omega$ and $\omega_{Fe}=0.3\cdot\omega$. This can be explained if we assume that massive stars explode in clusters and therefore they transfer more efficiently their energy into the ISM than isolated SNe such as Type Ia ones.

\item \textsl{Iron-enhanced differential wind} \citep{recchi2001}: the iron is expelled by galactic winds more efficiently than the $\alpha$-elements . This means that $\omega_{Fe}>\omega_{\alpha}$. This assumption is based on the fact that SNe Ia explode 
in a medium already heated and diluted by the previous activity of SNe II, as suggested by \citet{recchi2001}. However, \citet{recchi2004} have later shown that this conclusion is valid only for an isolated starburst.

\end{enumerate}

\par In Tab.(\ref{herculestable}), we report the predictions of our models; the first two columns specify each model in terms of its input parameters $M_{inf}$, $\nu$, and $\omega$. 
All the models predict a very small total gas mass at the present time, in agreement with the observations ($M_{gas,obs}<466\;M_{\odot}$, from \citealt{grcevich2009}). The actual total stellar mass of the galaxy 
($M_{\star,obs}=7.2^{+1.2}_{-1.1}\cdot10^{4}\;M_{\odot}$, from \citealt{martin2008}) is quite well reproduced by our models. However, we have to stress that such observed quantities remain still rather uncertain at the present time. 

\par All our models predict the MDF to peak at extremely low $[Fe/H]$ abundances. This is an effect both of the very low star formation efficiencies and of the very shallow potential well, which allows the galactic wind to develop earlier and, therefore, 
to diminish further the SFR in the subsequent evolution of the galaxy. Although we varied the $\nu$ parameter within a small range (from $\nu=0.002\;Gyr^{-1}$ up to $\nu=0.008\;Gyr^{-1}$), we obtained significant variations in the final predicted features 
of the galaxy.
\par As one can see in Tab.(\ref{herculestable}), diminishing the infall mass $M_{inf}$ will cause the MDF to peak at smaller $[Fe/H]$ values, once the other parameters are fixed. That is because the SFR has been assumed to be proportional to the gas mass 
inside the galaxy. In fact, the larger the gas mass available, the higher will be the star formation activity; so the chemical enrichment of the ISM will proceed at higher rates, shifting the MDF towards higher $[Fe/H]$.

\subsubsection{Hercules UfD: abundance ratios and interpretation} \label{hercules:abundance}

At extremely low $[Fe/H]$, it turns out that $[Ca/Fe]$ in the Hercules UfD steeply decreases from abundances above the solar values ($[Ca/Fe]=+0.32\pm0.22$ dex at $[Fe/H]=-3.10\pm0.16$ dex) down to $[Ca/Fe]=-0.51\pm-0,21$ dex at 
$[Fe/H]=-2.03\pm0.14$ dex \citep{aden2011}. In order to reproduce such a trend, confirmed also by successive observational spectroscopic studies of UfD stars \citep{gilmore2013,vargas2013}, we tested a formation scenario characterized by a very short 
infall time-scale ($\tau_{inf}=0.005$ Gyr) and very low star formation efficiencies ($0.003\;Gyr^{-1}\leq\nu\leq0.008\;Gyr^{-1}$), even lower than for dSphs and in agreement with previous studies \citep{salvadori2009}. The short infall time-scale gives 
rise to an enhanced SFR in the earliest stages of galaxy evolution, during which large amounts of $\alpha$-elements were likely to be ejected by Type II SN explosions into the ISM, whereas the extremely low star formation efficiency produces a very slow 
growth of the $[Fe/H]$; so when 
SNe Ia occur, the $[Fe/H]$ is still very low and the $[Ca/Fe]$ steeply decreases.

\par In Fig.(\ref{alphaherculesconfrontoventi}), we show the effect of the various wind implementations on the $[Ca/Fe]$ vs. $[Fe/H]$ diagram. An iron-enhanced \textsl{differential wind} is not able to reproduce the observed steep decrease in $[Ca/Fe]$ vs. 
$[Fe/H]$. Conversely, models with normal and $\alpha$-enhanced differential wind agree with the observations. By looking at this Figure, the three models predict the galactic wind to start at the same time. In fact, by fixing the $\tau_{inf}$ and 
$M_{inf}$ parameters, the time of the galactic wind onset depends only upon the features of the galaxy potential well (which affects the binding energy of the gas) and upon the SFH of the galaxy (which affects the computation of the thermal energy of the 
ISM). Therefore, the only parameter affecting the time of onset of the galactic wind is the $\nu$ parameter, whereas the $\omega$ parameter, as well as its particular implementation, comes into play only after the wind has started. An iron-enhanced 
differential 
wind predicts the $[Ca/Fe]$ to increase as $[Fe/H]$ grows, soon after the wind onset, at variance with observations. In fact, while SNe Ia enrich the ISM with large amounts of iron, causing the $[Fe/H]$ in the ISM to grow with time, the iron is also 
expelled by the galactic wind with higher efficiency than the $\alpha$-elements as the calcium. 
\par Since we do not still have evidences of galactic outflows enriched in $\alpha$-elements, we preferred as best models those with normal galactic wind.

\par In Fig.(\ref{calciumhercules}), we show the effect of changing the $\nu$ parameter on the $[Ca/Fe]$ vs. $[Fe/H]$ diagram, with $\omega=10$ (normal wind) and $M_{inf}=1.0\cdot10^{7}\;M_{\odot}$. The model which best reproduces the overall trend 
suggested by the dataset is characterized by $\nu=0.003\;Gyr^{-1}$ and $\omega=10$. In this Figure, we report also the observed $[Ca/Fe]$ vs. $[Fe/H]$ abundance pattern of Galactic halo stars. Although more data are necessary, Hercules UfD stars and 
Galactic halo stars display different trends of $[Ca/Fe]$ vs. $[Fe/H]$. In particular, halo stars do not share the fast decline in $[\alpha/Fe]$ exhibited by the stars in Hercules. This fact seems to rule out the hypothesis that UfDs would have been the 
building blocks of the stellar halo of the Milky Way. In fact, models (e.g. \citealt{brusadin2013}) suggest that the Galactic halo  have formed on a time-scale of $\approx0.2$ Gyr with a SF efficiency $\nu=0.2\;Gyr^{-1}$ and $\omega\simeq14$.

\begin{table*}
\begin{tabular}{c c c c c c c c c}
\hline
\multicolumn{9}{c}{\textbf{\normalsize Bo\"otes I: parameters of the models}}\\
\hline
\hline
$\nu$ & $\omega$ & $\tau_{inf}$ & SFH & $M_{inf}$ & $M_{DM}$ & $r_{L}$ & $S=\frac{r_{L}}{r_{DM}}$ & IMF \\
\hline 
$[Gyr^{-1}]$ & & [Gyr] & [Gyr] & $M_{\odot}$ & $M_{\odot}$ & $[pc]$ & & \\
\hline
\hline
$0.002/0.005/0.01/0.05$ & $10/15$ & $0.005$ & $0-4$ & $1.0/2.5/5.0\cdot10^{7}$ & $0.30\cdot10^{7}$ & $242$ & $0.2$ & \citet{salpeter1955} \\
\hline
\end{tabular}
\caption[bootesiinput]{ {\textit{Table:} we summarize here the specific characteristics of all the chemical evolution models performed for the Bo\"otes I UfD. \textit{Columns:} (1) star formation efficiency; (2) wind parameter; (3) infall time-scale; (4) 
period of major star formation activity \citep{dejong2008}; (5) total infall gas mass; (6) mass of the dark matter halo \citep{collins2013}; (7) effective radius of the \textsl{luminous} (baryonic) matter \citep{martin2008}; (8) ratio between the core radius of the dark 
matter halo and the effective radius of the \textsl{luminous} matter; (7) initial mass function.}}
\label{bootesiinput}
\end{table*}

\begin{table*}
\begin{tabular}{l c c | c c c c}
\hline
\multicolumn{7}{c}{\textbf{\normalsize Bo\"otes I chemical evolution models}}\\
\hline
\hline
\multicolumn{3}{l} {Input parameters}  & \multicolumn{4}{l}{Model predictions} \\
\multicolumn{1}{c}{} & $\nu$ & $\omega$ & $M_{\star,fin}$ & $M_{gas,fin}$  & $t_{wind}$ & $[Fe/H]_{peak}$ \\
\hline 
  & $[{Gyr}^{-1}]$ & & $M_{\odot}$ & $M_{\odot}$ & [Gyr] & dex \\
\hline
\hline
$M_{inf}=1.0\cdot10^{7}\;M_{\odot}$ & $0.002$ & $10$ & $0.14\cdot10^{5}$ & $0.35\cdot10^{1}$ & $1.17$ & $-2.99$ \\
           & $0.005$ & $10$ & $0.21\cdot10^{5}$ & $0.50\cdot10^{1}$ & $0.68$ & $-2.91$ \\
           & $0.01$ & $10$ & $0.29\cdot10^{5}$ & $0.14\cdot10^{1}$ & $0.45$ & $-2.81$ \\
           & $0.05$ & $10$ & $0.68\cdot10^{5}$ & $0.68\cdot10^{0}$ & $0.16$ & $-2.55$ \\
\hline
$M_{inf}=2.5\cdot10^{7}\;M_{\odot}$ & $0.002$ & $10$ & $0.58\cdot10^{5}$ & $0.90\cdot10^{1}$ & $2.01$ & $-2.68$ \\
           & $0.005$ & $10$ & $0.75\cdot10^{5}$ & $0.57\cdot10^{2}$ & $2.65$ & $-2.69$ \\
           & $0.01$ & $10$ & $0.11\cdot10^{6}$ & $0.39\cdot10^{2}$ & $0.76$ & $-2.51$ \\
           & $0.05$ & $10$ & $0.25\cdot10^{6}$ & $0.55\cdot10^{1}$ & $0.29$ & $-2.31$ \\
\hline
$M_{inf}=5.0\cdot10^{7}\;M_{\odot}$ & $0.002$ & $10$ & $0.19\cdot10^{6}$ & $0.19\cdot10^{2}$ & $3.21$ & $-2.39$ \\
           & $0.005$ & $10$ & $0.25\cdot10^{6}$ & $0.45\cdot10^{2}$ & $1.72$ & $-2.34$ \\
           & $0.01$ & $10$ & $0.33\cdot10^{6}$ & $0.89\cdot10^{2}$ & $1.12$ & $-2.28$ \\
           & $0.05$ & $10$ & $0.70\cdot10^{6}$ & $0.35\cdot10^{2}$ & $0.44$ & $-2.06$ \\
\hline
\end{tabular}
\caption[bootesitable]{{\textit{Table:} we reported for each model its main predictions. \textit{Columns:} (1) wind parameter; (2) star formation efficiency; (3) predicted actual total stellar mass; (4) predicted actual total gas mass; (5) time of the 
onset of the galactic wind; (6) peak of the stellar MDF predicted by the models.}}
\label{bootesitable}
\end{table*}

\subsubsection{Hercules UfD: MDF and interpretation} \label{hercules:mdf}

The observed stellar MDF \citep{aden2011} is built up on a very poor statistical sample, which consists of only $28$ RGB stars - the most luminous ones - previously identified and studied by \citet{aden2009b}. It turns out that most of the stars in the 
\citet{aden2011} sample resides in a very tight metallicity range, between $[Fe/H]=-3.1$ dex and $[Fe/H]=-2.9$ dex, with very low relative numbers of stars in the other bins towards both lower and higher $[Fe/H]$ abundances.  Conversely, our predicted 
stellar MDF is based upon a very large number of \textsl{artificial} stars determined from the assumed SFR and IMF. So, when we normalize the number of stars counted in a specific $[Fe/H]$-bin to the total number of stars, the height of the peak in the 
observed stellar MDF turns always out much higher than the more ``populated'' and dispersed MDF of our models. 

\par In Fig.(\ref{mdfhercules}), we have reported the comparisons between the observed MDF and our models with $M_{inf}=1.0\cdot10^{7}\;M_{\odot}$. By looking at the Figure, the only model matching the observations is characterized by a star formation 
efficiency $\nu=0.003\;Gyr^{-1}$; it is able to reproduce the $[Fe/H]$-position of the peak, as well as the width of the observed MDF, although it predicts a rather large relative number of stars in the extremely low $[Fe/H]$-wing, which spectroscopic 
observations have not yet been able to cover, because of the extremely low flux coming from those UfD member stars and the still relative low signal-to-noise ratio affecting the observations. By supposing that at extremely low $[Fe/H]$ abundances 
a hypothetical previous very massive population of stars might have enriched the medium with metals, the predicted relative number of stars with very low $[Fe/H]$ could be diminished, as it was similarly suggested in the past for solving the so-called 
\textsl{G-
dwarf problem} in the solar neighborhood. In this work, we only envisaged this hypothesis but we did not test its validity. Finally, by supposing that the chemical composition of the initial infall gas mass is not primordial but rather pre-enriched with 
metals by a previous population of stars, the observed enhanced $\alpha$-element abundances at extremely low $[Fe/H]$ might be explained. In fact, this might increase the initial metal content within the ISM and allow to build up that ``common 
\textsl{metallicity floor}'' to all the various different galaxy types, as suggested by \citet{ferrara2012}.
\par Our best model for Hercules requires $\nu=0.003\;Gyr^{-1}$, $\omega=10$ and $M_{inf}=1.0\cdot10^{7}\;M_{\odot}$.

\subsection{Chemical evolution of the Bo\"otes I UfD}

\begin{figure*}
 
    \includegraphics[width=14cm]{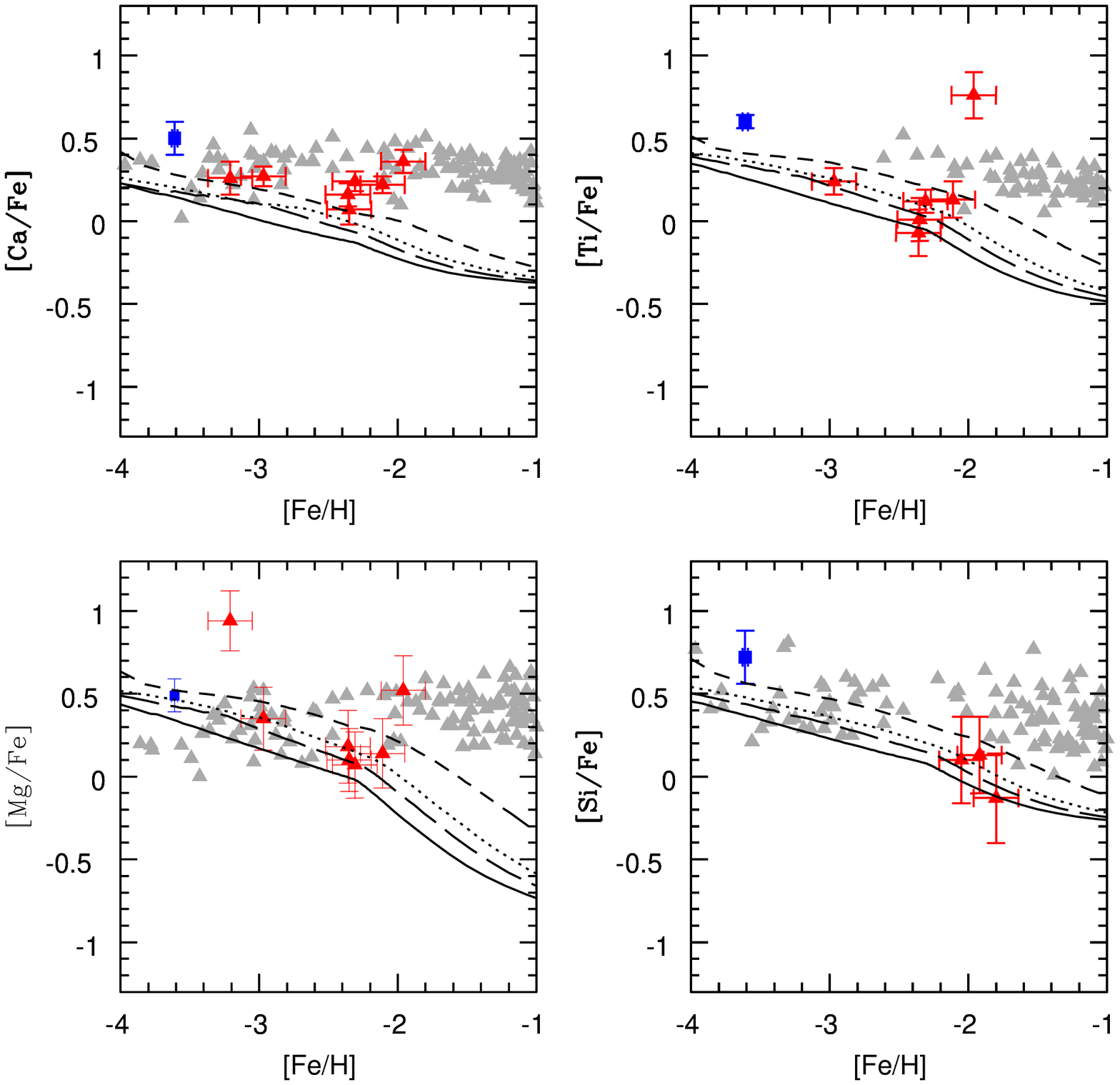} 
    \caption{
     In this Figure, we compare the $[\alpha/Fe]$ vs. $[Fe/H]$ abundance ratios for calcium, silicon, titanium and magnesium as observed in the Bo\"otes I UfD member stars with the trends predicted by our models with $\omega=10$ (normal wind) and 
     $M_{inf}=2.5\cdot10^{7}\;M_{\odot}$. The dataset consists of the data samples of \citet[blue squares]{norris2010a} and \citet[red triangles]{gilmore2013} (``NY'' analysis, from \citealt{norris2010b}). The model with $\nu=0.002\;Gyr^{-1}$ is given by the 
     solid line;  the model with $\nu=0.005\;Gyr^{-1}$ is given by the long dashed line, whereas the model with $\nu=0.01\;Gyr^{-1}$ is represented by the dotted line. Eventually, the model with $\nu=0.05\;Gyr^{-1}$ is represented by the dashed line. We 
     show also the abundance patterns of $[\alpha/Fe]$ vs. $[Fe/H]$ as observed in Galactic halo stars (grey triangles, data from \citealt{gratton2003,reddy2003,cayrel2004,reddy2006}).
    }
\label{alphabootesi}
\end{figure*}

We assumed for the Bo\"otes I UfD a dark matter halo with mass $M_{DM}=0.30\cdot10^{7}\;M_{\odot}$ \citep{collins2013}. The ratio between the core radius of the DM halo and the effective radius of the galaxy baryonic component has been assumed to be 
$S=\frac{r_{L}}{r_{DM}}=0.2$, where the effective radius of the baryonic (\textsl{luminous}) component of the galaxy is $r_{L}=242\;pc$ \citep{martin2008}. The SFH has been assumed to be made of only one episode lasting $4$ Gyr, from $0$ up to $4$ Gyr 
\citep{dejong2008}.

\par The infall mass of primordial gas from which the galaxy formed by accretion has been varied; we tested models with $M_{inf}=1.0$, $2.5$ and $5.0\cdot10^{7}\;M_{\odot}$. The infall time-scale of such initial reservoir of gas has been assumed to be 
very short, $\tau_{inf}=0.005$ Gyr. In Tab.(\ref{bootesiinput}), we reported the parameters of our Bo\"otes I chemical evolution models. All the models are characterized by a \textsl{normal} galactic wind.

\par In Tab.(\ref{bootesitable}), we report the main predictions of our chemical evolution models. In the first two columns, we listed the input parameters: $M_{inf}$ and $\nu$. From the Table, the lower the star formation efficiency $\nu$ - as well as 
the lower the infall mass $M_{inf}$ - the lower will be the $[Fe/H]$ at which the stellar MDF has its peak. Our models predict extremely low total gas content at the present time, as well as much lower total stellar mass than in dSph galaxies. This is 
again an effect both of the low initial infall gas mass and of the very low star formation efficiencies. The predicted final total gas masses are in agreement with the observed values ($M_{gas,obs}<86\;M_{\odot}$, from \citealt{grcevich2009}), as well as 
the predicted final total stellar masses agree quite well with the observed value $M_{\star,obs}=(6.7\pm0.6)\cdot10^{4}\;M_{\odot}$, from \citet{martin2008}, which is however very uncertain since it was inferred from a poor number of hypothetical stars belonging to 
the 
galaxy: $N_{\star}=324^{+28}_{-23}$, derived by means of a maximum likelihood algorithm applied to SDSS data. Furthermore, the observed total stellar mass strongly depends also on the IMF adopted and here we report the quantity derived by 
\citet{martin2008} when using the \citet{salpeter1955} IMF.

\subsubsection{Bo\"otes I UfD: abundance ratios and interpretation}

In order to reproduce the $[\alpha/Fe]$ vs. $[Fe/H]$ observed trends \citep{gilmore2013,norris2010a}, which are indeed rather uncertain, we adopted a very short infall time-scale - which enhances the SFR in the earliest stages of the galaxy evolution, 
giving rise to high $[\alpha/Fe]$ abundances at extremely low $[Fe/H]$ - coupled with very low star formation efficiencies - which cause the decrease in the $\alpha$-element abundances to start at very low $[Fe/H]$. We show our results in Fig.
(\ref{alphabootesi}), where we compare our theoretical curves with the observed dataset of the $[\alpha/Fe]$ vs. $[Fe/H]$ abundance patterns for calcium, silicon, titanium and magnesium. In this Figure, we compare also the chemical abundances of Bo\"otes 
I UfD member stars with those of Galactic halo stars.

\par The best agreement with the observed dataset is obtained by the model with $\nu=0.01\;Gyr^{-1}$, $\omega=10$ and $M_{inf}=2.5\cdot10^{7}\;M_{\odot}$. At extremely low $[Fe/H]$, the stars of Bo\"otes I display enhanced $[Ca/Fe]$ abundance ratios. Although our theoretical curves are able to reproduce the titanium, magnesium and silicon abundance patterns, there is still a discrepancy with calcium. That is probably due to the large uncertainty in the calcium yields from massive stars.

\subsubsection{Bo\"otes I UfD: MDF and interpretation}

The observed stellar MDF has been taken from \citet{lai2011}. Originally, their data sample was constituted by low-resolution spectra of $25$ stars, which then was extended to $41$ stars by adding non-overlapping stars from \citet{norris2010b} and 
\citet{feltzing2009}. This gave rise to an inhomogeneous collection of datasets, which is also a rather poor statistical sample. 
\par
\begin{figure}
    \includegraphics[width=9cm]{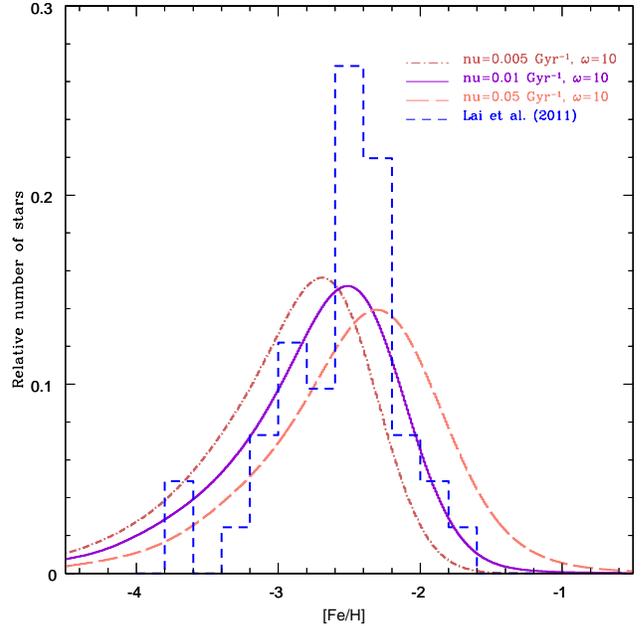} 
    \caption{
   In this Figure, we show the comparison between the observed stellar MDF of the Bo\"otes I UfD (data from \citealt{lai2011}) and the predictions of our models with $\omega=10$ (normal wind) and $M_{inf}=2.5\cdot10^{7}\;M_{\odot}$. Again, as in the case 
   of the Hercules UfD, the dataset is built up on a sample of only $41$ stars, highly concentrated in a very tight $[Fe/H]$ range. We are not able to reproduce the height of the observed-MDF peak. The model which best reproduces the $[Fe/H]_{peak}$ 
   abundance of the observed MDF turns out to have $\nu=0.01\;Gyr^{-1}$ and $\omega=10$.
    }
\label{mdfbootesi}
\end{figure}

 As in the case of the Hercules stellar MDF (see section \ref{hercules:mdf}), also here we are not able to reproduce the height of the observed MDF-peak. This is clearly illustrated in Fig.(\ref{mdfbootesi}),  where we also show that the only model 
able to reproduce the peak and the width of the observed MDF is characterized by $\nu=0.01\;Gyr^{-1}$, $\omega=10$ and $M_{inf}=2.5\cdot10^{7}\;M_{\odot}$.

\section{Conclusions} \label{section5}

\subsubsection{Summary on dSph chemical evolution}

In this paper, we have assumed that dSph galaxies form by accretion of an \textsl{infall mass} of $M_{inf}=1.0\cdot10^{8}\;M_{\odot}$ with primordial chemical composition. The infall rate obeys to a decaying exponential law with an \textsl{infall time-
scale} $\tau_{inf}=0.5$ Gyr. The mass content and the core radius of the DM halo have been taken from the observations. In particular, here we modeled Sculptor and Carina and our results can be summarized as follows.

\begin{enumerate}

\item Only by assuming low star formation efficiencies ($\nu=0.05-0.5\;Gyr^{-1}$) relative to the Milky Way ($\nu\sim1\;Gyr^{-1}$ in the solar vicinity, \citealt{chiappini1997}), we have been able to explain the decline in the $[\alpha/Fe]$ abundance 
ratios observed at very low $[Fe/H]$. In fact, the lower the $\nu$ parameter, the lower is the amount of iron produced by Type II SNe; this  causes the $[Fe/H]$ of the ISM, at which Type Ia SNe become predominant in the iron pollution, to be lower. This 
precisely determines the decrease in the $[\alpha/Fe]$ at very low $[Fe/H]$. This is due to the fact that most of $\alpha$-elements originates in Type II SNe whereas most of Fe originates in Type Ia SNe exploding on a large range of times (time-delay 
model).

\item Low star formation efficiencies increase the star formation time-scales, giving rise to stellar MDFs peaked towards very low $[Fe/H]$, in very good agreement with the observations indicating that dSphs are very metal-poor stellar systems.

\item Galactic winds - driven by SN explosions - play a relevant role in the dSph chemical evolution. We adopted for dSphs a \textsl{normal} wind, i.e. a wind with an efficiency equal for all the chemical elements. The net effect of the galactic wind is 
to diminish further the SFR and cause the $[\alpha/Fe]$ to follow a steeper decline as $[Fe/H]$ increases. Thanks to the strong outflow rates, our models predict very low final total gas masses, according to observations.

\item Only by adopting star formation episodes occurring far beyond the reionization epoch, our models were able to build up final total stellar masses of the same order of magnitude as the observed ones ($M_{star,obs}\sim10^{6}\;M_{\odot}$), and they 
could also reproduce the other observed features of dSph galaxies. Our assumptions were based on the results of the CMD-fitting analysis, which have been able, in the past few years, to derive the SFH of many dwarf galaxies of the Local Group.

\item We adopted the \citet{salpeter1955} IMF for all the models. In a recent paper, \citet{mcwilliam2013} suggested that to explain the abundance data in Sagittarius an IMF with a lower fraction of massive stars than a Salpeter-like IMF should be preferred. However, by testing an IMF with a steeper slope in the massive star range, we found a neglibible difference in our results.

\item Our best models are in good agreement with the previous works \citep{lanfranchi2004,lanfranchi2006b} and confirm their goodness. Our best model for Sculptor requires $\nu=0.05\;Gyr^{-1}$ and $\omega=10$, whereas our best model for Carina requires 
$\nu=0.2\;Gyr^{-1}$ and $\omega=10$.

\end{enumerate}

\subsubsection{Summary on UfD chemical evolution}

We modeled Hercules and Bo\"otes I, by adopting the same numerical code of dSphs but with different characteristic input parameters. In what follows, we summarize the main features of our UfD chemical evolution models:

\begin{enumerate}

\item Since UfDs are known today to be the faintest and most DM dominated galaxies, we adopted \textsl{infall masses} $M_{inf}=(1.0-5.0)\cdot10^{7}\;M_{\odot}$, which are lower than the ones assumed for dSphs and supported by observations. Such initial 
reservoir of gas, with primordial chemical composition, was accreted in the galaxy DM halo on a very short typical time-scale ($\tau_{inf}=0.005$ Gyr).

\item Our best model for Hercules requires $\nu=0.003\;Gyr^{-1}$, $\omega=10$, $M_{inf}=1.0\cdot10^{7}\;M_{\odot}$ and $\tau_{inf}=0.005$ Gyr, whereas our best model for Bo\"otes I requires $\nu=0.01\;Gyr^{-1}$, $\omega=10$, 
$M_{inf}=2.5\cdot10^{7}\;M_{\odot}$ and $\tau_{inf}=0.005$ Gyr.

\item The $[\alpha/Fe]$ abundance ratios are observed to steeply decrease at very low $[Fe/H]$. This is clearly a signature of very low star formation efficiencies ($\nu=0.001-0.01\;Gyr^{-1}$), even lower than for the dSph ones 
($\nu\approx0.05-0.5\;Gyr^{-1}$). In fact, Type Ia SNe start to dominate in the iron pollution of the ISM when the $[Fe/H]$ of the ISM is still very low. This result is in agreement with the previous studies of UfD galaxies, such as \citet{salvadori2009}

\item The very low adopted star formation efficiencies cause the stellar MDFs to be peaked at very low $[Fe/H]$, in very good agreement with observations.

\item From the CMD-fitting analysis, as well as from the study of the pulsation and metallicity properties of their variable stars, UfDs host very old stellar populations, with ages $>10-12$ Gyr. So, we assumed them to have undergone star formation only 
in the first gigayears of their chemical evolution history. This fact, coupled with the very low star formation efficiencies, causes the building up of very low final total stellar masses $M_{\star,fin}\sim10^{4}-10^{5}\;M_{\odot}$, which are much lower 
than the observed dSph ones, and in agreement with observational values.

\item The actual total gas masses in UfDs are negligible or even undetected. This is a clear evidence of intense galactic winds, which are efficient in the gas removal from the galaxy potential well and started very soon in the galaxy evolution.  We 
tested models with \textsl{normal wind}, $\alpha$-enhanced \textsl{differential wind}, and iron-enhanced \textsl{differential wind}. All such models agree in predicting negligible final total gas masses but each of them has a different effect in the 
$[\alpha/Fe]$ vs. $[Fe/H]$ diagram. 

\item Both \textsl{normal} and $\alpha$-enhanced galactic winds predict a trend of the $[\alpha/Fe]$ vs. $[Fe/H]$ diagram which well agrees with observations. On the other hand, an iron-enhanced \textsl{differential wind} predicts the $[\alpha/Fe]$ to 
increase as $[Fe/H]$ grows once the galactic wind has started, at variance with observations. We suggest the galactic wind to be normal with an efficiency $\omega=10$, which is the same value adopted in this work for reproducing the main properties of 
dSph galaxies.

\item Our chemical evolution models have been able to reproduce, at the same time and reasonably well, all the observed features of the two UfD galaxies studied, such as the abundance ratio patterns of the $\alpha$-elements, the stellar MDF, and the total 
stellar and gas masses at the present time. From the main characteristics of our chemical evolution models and from the observed trend of the UfD chemical abundances, we suggest that the hypothesis that UfDs would have been the survived ``building blocks'' 
of the Milky Way halo is unlikely, since the two seem to have undergone very different galactic chemical enrichment histories. In particular, the Galactic halo abundance pattern suggests more vigorous star formation and galactic wind, together with a 
longer time-scale of gas accretion (see \citealt{brusadin2013}). However, more data on UfDs are necessary before drawing firm conclusions.

\end{enumerate}

\section*{Acknowledgments}
F.M. acknowledges financial support from PRIN-MIUR~2010-2011 project ``The Chemical and Dynamical Evolution of the Milky Way and Local Group Galaxies'', prot.~2010LY5N2T. F.V. and F.M. thank M. Nonino, A. Koch and D. Romano for useful suggestions and interesting discussions. We thank an anonymous referee for his/her suggestions.

\bsp

\label{lastpage}

\end{document}